\documentclass[10pt,journal,compsoc]{IEEEtran}

\usepackage[nocompress]{cite}
\usepackage{
  amsmath,
  amssymb,
  amsthm,
  thmtools,
  graphicx,
  xcolor,
  multirow,
  url,
}
\usepackage[linkcolor=black,colorlinks=true,citecolor=teal,urlcolor=cyan,hypertexnames=false]{hyperref}
\usepackage[capitalise]{cleveref}
\usepackage{lipsum} 
\usepackage[switch]{lineno}
\usepackage[raggedrightboxes]{ragged2e}
\newif\ifpeerreview
\peerreviewfalse

\graphicspath{{images/}}

\newcommand{\paperID}{4}

\usepackage[numbers,sort&compress]{natbib}

\usepackage{soul}

\newcommand{\notecolor}{violet}
\newcommand{\delcolor}{lightgray}

\newcommand{\change}[1]{#1}

\newcommand{\RR}{\mathbb{R}}

\newcommand{\norm}[1]{\lVert #1 \rVert}

\newcommand{\sym}[1]{\mathtt{#1}} 
\renewcommand{\vec}[1]{\mathbf{#1}}
\newcommand{\mat}[1]{\vec{#1}}

\DeclareMathOperator*{\argmin}{argmin}
\IEEEoverridecommandlockouts

\title{Spectral Sensitivity Estimation Without a Camera}
\author{Grigory~Solomatov,~\IEEEmembership{Member,~IEEE,}
        and~Derya~Akkaynak,~\IEEEmembership{Member,~IEEE}
\IEEEcompsocitemizethanks{\IEEEcompsocthanksitem G. Solomatov and D. Akkaynak are with the Hatter Department
of Marine Technologies, Leon H. Charney School of Marine Sciences, University of Haifa, and the Interuniversity Institute for Marine Sciences in Eilat \protect\\
E-mail: gsolomat@campus.haifa.ac.il}
}

\begin{document}

\IEEEtitleabstractindextext{%
\begin{abstract}
A number of problems in computer vision and related fields would be mitigated if camera spectral sensitivities were known. As consumer cameras are not designed for high-precision visual tasks, manufacturers do not disclose spectral sensitivities. Their estimation requires a costly optical setup, which triggered researchers to come up with numerous indirect methods that aim to lower cost and complexity by using color targets. However, the use of color targets gives rise to new complications that make the estimation more difficult, and consequently, there currently exists no simple, low-cost, robust \emph{go-to} method for spectral sensitivity estimation that non-specialized research labs can adopt. Furthermore, even if not limited by hardware or cost, researchers frequently work with imagery from multiple cameras that they do not have in their possession. To provide a practical solution to this problem, we propose a framework for spectral sensitivity estimation that not only does not require \emph{any} hardware (including a color target), but also does not require physical access to the camera itself. Similar to other work, we formulate an optimization problem that minimizes a two-term objective function: a camera-specific term from a system of equations, and a universal term that bounds the solution space. Different than other work, we utilize publicly available high-quality calibration data to construct both terms. We use the colorimetric mapping matrices provided by the Adobe DNG Converter to formulate the camera-specific system of equations, and constrain the solutions using an autoencoder trained on a database of ground-truth curves. On average, we achieve reconstruction errors as low as those that can arise due to manufacturing imperfections between two copies of the same camera. We provide our code and predicted sensitivities for $1,000+$ cameras at https://github.com/COLOR-Lab-Eilat/Spectral-sensitivity-estimation, and discuss which tasks can become trivial when camera responses are available. 
\end{abstract}

\begin{IEEEkeywords} 
Adobe DNG SDK, colorimetric mapping, color constancy, reflectance recovery, illumination estimation
\end{IEEEkeywords}
}

\maketitle
\thispagestyle{empty} 

\IEEEraisesectionheading{
  \section{Introduction}\label{sec:introduction}}
  \IEEEPARstart{C}{onsumer} digital cameras (i.e., point-and-shoot, DSLR, mirrorless cameras and smartphones) are not tools designed for scientific imaging, i.e., they are not scientific light-measuring devices~\cite{burggraaff2019standardized,akkaynak2014use,kim2012new}. Yet their outputs\textemdash digital photos and videos\textemdash constitute major sources of data for image processing, colorimetry, computational photography, computer vision, and machine learning. Typically, research utilizing consumer camera imagery focuses on the development of filters or algorithms that alter the visual appearance of images~\cite{paris2009fast, guo2020zero,zhu2017unpaired,bychkovsky2011learning}, or on the understanding of scene content~\cite{deng2009imagenet,lowe1999object,xiao2010sun} and structure~\cite{mildenhall2021nerf,lee2009geometric} with downstream goals like recognizing, tracking, or counting objects. For many of these goals, successfully recovering scene reflectance and/or illumination is key (and often the main goal itself), but these tasks are complicated by the fact that consumer cameras do not capture colors in a standardized way~\cite{nguyen2014raw,chakrabarti2009empirical}.

\begin{figure}[!t]
\centering
\includegraphics[width = \columnwidth]{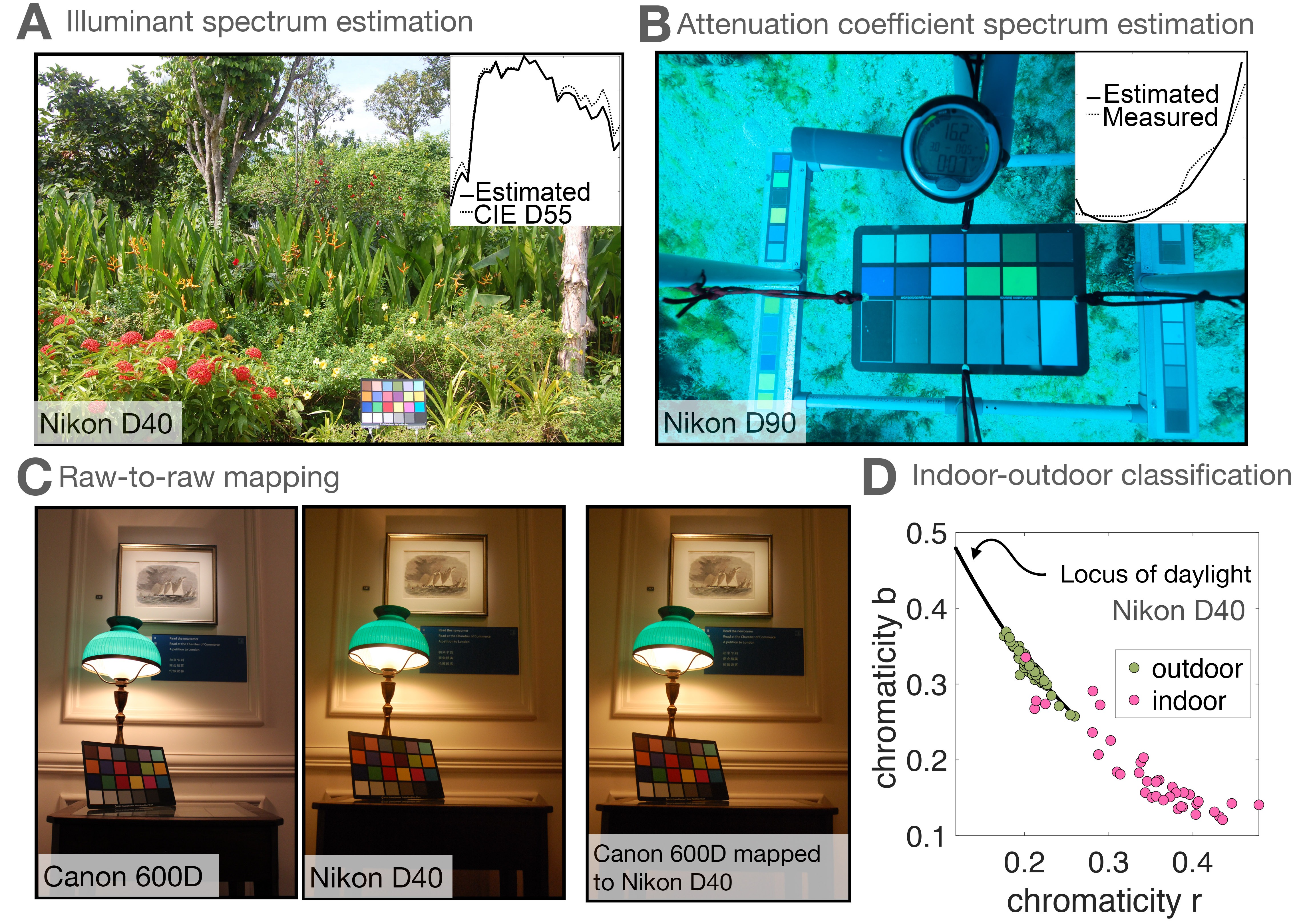}
\caption{\label{fig:fig1}Examples of tasks that can become possible or easier with known camera response. \textbf{A)} The spectrum of a uniformly lit outdoor scene can be recovered using a color chart. Image from~\cite{nguyen2014raw}. \textbf{B)} Underwater, when the imaging distance is small, the diffuse downwelling attenuation coefficient can be recovered using a color chart. Data from~\cite{akkaynak2017space}. \textbf{C)} When mapping one camera's response to another, the tedious pairwise imaging and alignment~\cite{nguyen2014raw} or the costly supervised learning steps~\cite{afifi2021semi} can be entirely omitted. Data from~\cite{cheng2014illuminant}. \textbf{D)} The locus of a camera's daylight chromaticities can be used to classify whether the image was taken indoors or outdoors. Images from~\cite{cheng2014illuminant}. \change{These results were obtained by implementing previously published methods in a way that incorporates the spectral sensitivities of the relevant camera. We did not have physical access to any of the cameras used; see Supplementary Materials for details.}}\vspace{-10pt}
\end{figure}

  Each camera captures colors in its own, device-specific color space. This is because manufacturers make different design choices for each sensor, which materialize within the limitations inherent to silicon-based semiconductor technology~\cite{bhandari2022computational}. As a result, each make and model of camera is characterized by its unique set of spectral sensitivities that define how that camera responds to light. Effectively, this means that different cameras will register a different set of \texttt{RGB} values for a given scene\textemdash even if the scene is imaged under identical conditions~\cite{nguyen2014raw,karaimer2018improving}. If the camera's color bias could be eliminated, the recovery of scene reflectance (e.g., for material property estimation~\cite{munoz2011bssrdf,debevec2004estimating,yu1999inverse}, spectral super-resolution~\cite{oh2016yourself,galliani2017learned}, etc.) and/or scene illumination (e.g., for computational color constancy~\cite{barnard2002camera,cheng2014illuminant,lo2021clcc}, image relighting~\cite{el2021ntire}, data augmentation~\cite{afifi2021cross,afifi2021cie,lou2015color} etc.) could be achieved simpler and faster, and with more consistent results across cameras. Figure~\ref{fig:fig1} gives some examples of tasks that could become possible or easier if camera spectral sensitivities were known.

  Unfortunately, however, manufacturers of consumer cameras rarely make spectral sensitivity curves available, and the burden falls on the researchers to derive them. Their derivation (detailed in~\cref{sec:background}) requires the stimulation of the camera's sensor with monochromatic light across the entire wavelength range to which it is sensitive. Such light can be obtained through the use of a monochromator or a spectrally-tuneable light source, but these optical instruments are generally too specialized and expensive to be found in most research labs. A seemingly lower cost and complexity alternative is to use photos of calibrated color targets, but with this indirect option comes a hidden cost:  the implicit requirement that the illumination under which the color target is imaged is \emph{also} known. In practice, this either means that the light source is calibrated, or its spectrum is measured with a spectrophotometer; necessitating, again, that the user has access to non-standard hardware (\cref{tab:survey}). 
  
  Even when the light spectrum is known, photos of color charts do not guarantee a ``good enough'' estimate of curves because there are too many unknowns in camera spectral sensitivities and not enough color chart patches whose reflectances are linearly independent~\cite{alsam2002recovering,hardeberg2002spectral,parkkinen1989characteristic}. The ill-posed nature of the problem introduces new challenges due to the need to constrain the solutions to a ``realistic'' space (\cref{tab:survey}). Furthermore, the commonly used image formation model does not take into account the optoelectronics of the camera~\cite{martinez2002calculation}, making it difficult to relate the captured \texttt{RGB} values to the predicted ones. Finally, even if not limited by hardware or cost, it is not uncommon that researchers work with imagery captured by cameras that they actually do not have in their possession. Consequently, the missing knowledge regarding the spectral responses of consumer cameras hampers the otherwise fruitful utilization of a consumer electronic device in academic research.

Our own research in underwater computer vision has been held back because we did not have the spectral sensitivities of the ever-evolving set of consumer cameras that produce the imagery we work with, and this motivated us to search for a solution that did \underline{not} require: 1) any specialized hardware, 2) a color chart photo taken by the camera, or 3) physical access to the camera itself. Unlike how it may seem at first glance, this is not an impossible task\textemdash one simply has to be aware of the wealth of publicly available data regarding colorimetric mapping of consumer digital cameras. Here, we describe a method that satisfies these requirements with reconstruction errors that are, on average, as low as those that can arise due to manufacturing imperfections between two copies of the same camera.
 
Our key observation is the following: Among the tags specific to Adobe's Digital Negative (DNG) format~\cite{AdobeDNGSpec}, are color transformation matrices prefixed \emph{ColorMatrix}. These 3x3 matrices map camera-specific colors to the standard CIE XYZ color space under a known illuminant. Calibrated by the manufacturer, Adobe,  or third-party companies and appended to image metadata by the Adobe DNG Converter, each DNG file includes at least one \emph{ColorMatrix} tag, whereas frequently, two matrices per camera can be found\footnote{Since the release of DNG version 1.6.0.0, a third color matrix is also specified for some cameras~\cite{AdobeDNGSpec}.}.  In the common case when two matrices are available per camera, they are generally calibrated under illuminant A (tungsten, CCT$\approx$2856K) and D65 (daylight, CCT$\approx$6504K).
We use these matrices to construct a linear system of equations which contains the sought spectral sensitivities in its solution space; and this solution space can be further constrained using an autoencoder trained on database of ground-truth spectral sensitivities from the literature. Using this novel approach we make the following contributions:
  \begin{enumerate}
  \item a zero-cost method with which to obtain reasonable estimates of camera spectral sensitivities, and
  \item predicted spectral sensitivities of more than $1,000$ makes and models of cameras that are currently in the market.
  \end{enumerate}
  In addition, we provide a thorough survey of indirect spectral sensitivity estimation methods that have evolved over the last three decades, corresponding with the advent of consumer cameras as prevalent research tools.

\section{Background and related work}
\label{sec:background}   
    
The gold standard of spectral sensitivity estimation is the use of a monochromator setup in an optically sealed environment~\cite{darrodi2015reference,macdonald2002colour,vora1997digital,martinez2002calculation}. A monochromator generates light in narrowband segments, which is then shone directly onto the sensor being calibrated or photographed from a calibration tile with uniform reflectance. At the same time as the light shines on the sensor, its spectrum is measured using a spectrophotometer, whose signal is then used as the reference for recovering the camera's response from its captured \texttt{RGB} values. Such a setup is described in detail in~\cite{darrodi2015reference}. 
 
While a monochromator setup provides a \emph{direct} way of measuring spectral sensitivity curves and yields results with high accuracy, it consists of specialized hardware that can be prohibitively expensive, or simply inaccessible, for many research labs. Thus, over the last 30 years since consumer digital cameras became commonplace tools in research labs, numerous \emph{indirect} methods, which aim to estimate a camera's spectral response at a lower cost or complexity than the monochromator method, have been proposed (\cref{tab:survey}).

\newcommand{\tabsurveyrow}[6]{#1 & #2 & #4 & #3 & #5 & #6 \\}
\newcommand{\tabsurveyrowopts}[6]{#1 | #2 | #4 | #3 | #5 | #6}
\begin{table*}[t!]
\renewcommand{\arraystretch}{1.3}
\caption{\label{tab:survey} A summary of indirect spectral sensitivity estimation methods developed for digital cameras in the last 30 years and their requirements. }\centering
\begin{tabular}{\tabsurveyrowopts{p{1.5in}}{p{1.7in}}{p{0.57in}}{p{1.4in}}{p{0.5in}}{p{0.4in}}}

\hline
\tabsurveyrow{\textbf{Reference} }{ \textbf{Approach} }{ \textbf{\footnotesize{Camera required?}} }{ \textbf{Specialized hardware required?} }{ \textbf{Chart \mbox{required?}} }{ \textbf{Light known?} }
\hline\hline
\tabsurveyrow{Hubel et al.~(1994)\cite{hubel1994comparison}  }{ Linear regression, rank deficient pseudoinverse, Wiener method }{ Yes }{ Tungsten light, color filters, spectrophotometer }{ Yes }{ Yes  }
\tabsurveyrow{Hardeberg et al.~(1998)\cite{hardeberg1998spectral} }{ Pseudoinverse, rank deficient pseudoinverse }{ Yes }{ No }{ Yes }{ Yes }
\tabsurveyrow{Finlayson et al.~(1998)\cite{finlayson1998recovering} }{ Constrained linear regression }{ Yes }{ No }{ Yes }{ Yes  }
\tabsurveyrow{Barnard \& Funt~(1999)\cite{barnard1999camera} }{ Constrained linear regression }{  Yes }{ Multiple illuminants, color filters, spectroradiometer }{ Yes }{ Yes }
\tabsurveyrow{Dyas~(2000)\cite{dyas2000robust} }{ SVD + regularization }{ Yes }{ No }{ Yes }{ Yes}
\tabsurveyrow{Carvalho et al.~(2001)\cite{carvalho2001learning} }{ Learning with generalized cross validation function }{ Yes }{ No }{ Yes }{ Yes  }
\tabsurveyrow{Barnard \& Funt~(2002)\cite{barnard2002camera} }{ Constrained linear regression }{ Yes }{ Multiple illuminants, color filters, spectroradiometer }{ Yes }{Yes  }
\tabsurveyrow{Alsam \& Finlayson~(2002)\cite{alsam2002recovering} }{ Feasible set }{ Yes }{ No }{ Yes }{ Yes  }
\tabsurveyrow{Quan et al.~(2003)\cite{quan2003comparative} }{ Iterative multiscale bases }{ Yes }{ No }{ Yes }{ Yes  }
\tabsurveyrow{DiCarlo et al.~(2004)\cite{dicarlo2004emissive} }{ Emissive target }{ Yes }{ LEDs, opto-electronics }{ Emissive }{ Yes  }
\tabsurveyrow{Ebner~(2007)\cite{ebner2007estimating} }{ Evolution strategy }{ Yes }{ Spectrometer }{ Yes }{ Yes}
\tabsurveyrow{Mauer \& Wueller~(2009)\cite{mauer2009measuring} }{ Narrowband interference filters }{ Yes }{ Filters and mount }{ No }{ Yes}
\tabsurveyrow{Zhao et al.(2009)\cite{zhao2009estimating} }{ Optimum basis functions }{ Yes }{ Spectrometer  }{ Yes }{ Yes}
\tabsurveyrow{Rump et al.~(2011)\cite{rump2011practical} }{ Color specularity \& light nonuniformity + optimization }{ Yes }{ Spectrometer }{ Yes }{ Yes  }
\tabsurveyrow{Pike~(2011)\cite{pike2011using} }{ Parameterization of\cite{finlayson1998recovering} + UV }{ Yes }{ Spectrometer, UV target }{ Yes }{ Yes  }
\tabsurveyrow{Bedarkar~(2011)\cite{bedarkar2011camera} }{ Optimization  }{ Yes }{ Blackbody furnace }{ No }{ \scriptsize{Planck's law  }}
\tabsurveyrow{Han et al.~(2012)\cite{han2012camera} }{ Fourier bases }{ Yes }{ Fluorescent chart }{ Yes }{ No  }
\tabsurveyrow{Jiang et al.~(2013)\cite{jiang2013space} }{ PCA }{ Yes }{ None }{ Yes }{ \scriptsize{Yes/\mbox{daylight}}}
\tabsurveyrow{Prasad et al.~(2013)\cite{prasad2013quick} }{ Optimization }{ Yes }{ Multiple illuminants }{ Yes }{ No }
\tabsurveyrow{Kawakami et al.~(2013)\cite{kawakami2013camera} }{ Sky turbidity, need clear sky \& camera-sun direction + PCA }{ Yes }{ No }{ No }{ No }
\tabsurveyrow{Bongiorno et al.~(2013)\cite{bongiorno2013spectral} }{ Linear variable edge filter (LVEF) }{ Yes }{ LVEF, smooth light, spectrometer, Spectralon }{ No }{ Yes }
\tabsurveyrow{Huynh \& Robles-Kelly~(2014)\cite{huynh2014recovery} }{ Optimization }{ Yes }{ No }{ Yes }{ No }
\tabsurveyrow{Macdonald~(2015)~\cite{macdonald2015determining} }{ Emissive target + transmission filters  }{ Yes }{ Spectrometer, emissive target, color filters }{ Emissive }{ Yes }
\tabsurveyrow{Bartczak et al.~(2015)~\cite{bartczak2015led} }{ Emissive target }{ Yes }{ Spectrometer, spectrally-tuneable light, diffuser }{ Emissive }{ Yes }
\tabsurveyrow{Finlayson et al.~(2016)\cite{finlayson2016rank} }{ Rank-based + sine bases }{ Yes }{ No }{ Yes }{ Yes }
\tabsurveyrow{Manakov~(2016)\cite{manakov2016evaluation} }{ Band-pass filters }{ Yes }{ Color filters, integrating sphere, spectrometer }{ No }{ Yes }
\tabsurveyrow{Zhang et al.~(2017)\cite{zhang2017camera} }{ Regularized least squares }{ Yes }{ Hyperspectral sensor }{ No }{ No }
\tabsurveyrow{Chaji et al.~(2018)\cite{chaji2018estimation}  }{ Neural network }{ Yes }{ No }{ Yes }{ \scriptsize{Daylight}}
\tabsurveyrow{Karge~(2018)\cite{karge2018using}}{ Diffraction image + least squares  }{ Yes }{ Grating, light, radiometer }{ Yes }{ Yes}
\tabsurveyrow{Burggraaff et al.~(2019)\cite{burggraaff2019standardized} }{ Pocket spectrometer + solar irradiance model  }{ Yes }{ Spectrometer, halogen lamp }{ No }{ Yes }
\tabsurveyrow{Zhu et al.~(2020)\cite{zhu2020spectral} }{ Orthogonal test + optimization }{ Yes }{ Spectral irradiance colorimeter, LCD display }{ Yes }{ Yes }
\tabsurveyrow{Toivonen \& Klami~(2020)~\cite{toivonen2020practical}  }{ Diffraction imaging + optimization }{ Yes }{ Diffuser, diffraction grating, color filters }{ Yes }{ Yes }
\tabsurveyrow{Ji et al.~(2021)~\cite{ji2021compressive} }{ Compressive sensing }{ Yes }{ Xenon-arc lamp }{ Yes }{ Yes }
\tabsurveyrow{Tominaga et al.~(2021)\cite{tominaga2021measurement} }{ PCA \& color patches }{ Yes }{ Incandescent light, Munsell white paper, Spectralon }{ Yes }{ Yes  }
\tabsurveyrow{Ma et al.~(2021)\cite{ma2021recovery} }{ Constrained least squares + sine bases }{ Yes }{ No }{ Yes }{ Yes }
\tabsurveyrow{Xu et al.~(2022)\cite{xu2022rank} }{ Rank-based }{Yes }{ Multiple illuminants }{ Yes }{ Yes }
\tabsurveyrow{Fan \& Ronnier~(2023)\cite{fan2023camera}}{ Emissive target + PCA }{ Yes }{ OLED display, radiometer }{ Emissive }{ Yes }
\tabsurveyrow{Ours}{ Color matrices + autoencoder }{ \textbf{No} }{ \textbf{No} }{ \textbf{No} }{ \textbf{No} }
\hline
\end{tabular}
\end{table*}

Despite the apparent diversity in techniques for indirect spectral sensitivity reconstruction, their core strategy rarely deviates from the same underlying principles, and our method is no different. At the outset, the sought sensitivity functions are described as a solution to a system of equations, typically linear or quadratic, which generally turns out to be underdetermined in the sense of admitting infinitely many solutions. Additional constraints such as non-negativity, smoothness and unimodality that are then assumed to hold for a broader family of spectral sensitivities are consequently employed with the aim of discarding spurious and ``unrealistic'' candidate solutions. Finally, the approximate sensitivity functions are obtained by solving an optimization problem whose objective function is comprised of two terms: one that originates from the system of equations and is specific to the camera under consideration, and one that corresponds to additional assumptions made about the space of functions that presumably contains the sought sensitivities. We will refer to these terms as the \emph{specific term} and the \emph{universal term}, respectively; the former depending on the data available from the experimental setup, and the latter depending on assumptions that may or may not hold true, since nobody really knows what spectral sensitivities of future cameras could look like (although we can probably expect them to become more colorimetric).

There is an inherent interplay between these two terms in the sense that the universal term compensates for the lack of information in the specific term. Indeed, using a monochromator yields nearly perfect information for the specific term, rendering the universal term unnecessary. Conversely, if no experimental data is available from the camera under consideration, then the best approximation one could hope for would be some kind of ``averaging'' of all known spectral sensitivities, which constitutes an extremely strong assumption on the space of candidate functions.\footnote{\change{The particular kind of ``averaging'' would depend on the targeted error metric. We computed such an ``average'' with respect to the \emph{relative full-scale error} (see \cref{sec:results}), minimizing the average of this error across our dataset. This yielded $21\%$ median error, which is much worse than the $8\%$ achieved by our method. This clearly demonstrates that color matrices do contain a lot of useful information about spectral sensitivities.}} With this trade-off in mind, we set out to devise a method that achieves the following:
\begin{enumerate}
\item the experimental setup required by the specific term is minimal; and
\item the universal term is as unassuming as possible while ensuring 1).
\end{enumerate}

We propose a method that attains 1) by constructing the specific term using publicly calibration matrices, which does not even require the user to own the camera in question, while 2) is achieved by training an autoencoder on a database of known spectral sensitivities. 

\subsection{The image formation model and its challenges}
\label{subsec:img-formation}

In this subsection, we discuss the physical image formation model that gives spectral sensitivity functions their meaning; the discretized version of this model is by far the most common choice for constructing the specific term in the optimization problem used for spectral sensitivity reconstruction. 

For any wavelength $\lambda$, we denote the  sensitivity of color channel $k \in \{\sym{r}, \sym{g}, \sym{b}\}$ in a device $d$ to light of that wavelength by $S_{d}^{(k)}(\lambda)$. If a target object with reflectance $R(\lambda)$ is photographed by $d$ under some illuminant $L(\lambda)$, then the resulting pixel intensity for channel $k$ is given by

\begin{equation}
  \label{eq:image-model-cont}
  I_{d}^{(k)} = \int_{0}^{\infty}R(\lambda)L(\lambda)S_{d}^{(k)}(\lambda) d\lambda \ .
\end{equation}
The sensitivity $S_{d}^{(k)}$ is typically zero outside of the visible spectrum $400$\,--\,$700\text{nm}$, in which case the above integral only needs to be considered on this interval. Doing so makes it possible to discretize the above image formation model over $n$ uniformly spaced values of $\lambda$ as follows:

\begin{equation}
  \label{eq:image-model-disc}
  I_{d}^{(k)} \approx \sum_{i=1}^n R(\lambda_i)L(\lambda_i)S_{d}^{(k)}(\lambda_i) \Delta \lambda \ ,
\end{equation}
where
\begin{align*}
  \Delta \lambda &= \frac{700 - 400}{n-1}\text{nm} = \frac{300}{n-1}\text{nm} \quad \text{and} \\
  \lambda_i &= 400\text{nm} + (i-1) \Delta \lambda \quad \text{for } i=1,\dots,n \ .
\end{align*}
Rewriting \eqref{eq:image-model-disc} in matrix form yields:
\begin{equation}
  \label{eq:image-model-mat}
  \mat{I}_d \approx \mat{R}\mat{L}\mat{S}_d \ , 
\end{equation}
where
\begin{align*}
  \mat{I}_d &=
              \begin{bmatrix}
                I_d^{(\sym{r})} & I_d^{(\sym{g})} & I_d^{(\sym{b})}
              \end{bmatrix} \in \RR^{1 \times 3} \ , \\
  \mat{R} &=
            \begin{bmatrix}
              R(\lambda_1) & \cdots & R(\lambda_n)
            \end{bmatrix} \in \RR^{1 \times n} \ , \\
  \mat{S}_d &= \begin{bmatrix}
                 S_d^{(\sym{r})}(\lambda_1) & S_d^{(\sym{g})}(\lambda_1) & S_d^{(\sym{b})}(\lambda_1) \\
                 \vdots & \vdots & \vdots \\
                 S_d^{(\sym{r})}(\lambda_n) & S_d^{(\sym{g})}(\lambda_n) & S_d^{(\sym{b})}(\lambda_n)
               \end{bmatrix} \in \RR^{n \times 3} \ \ \text{and} \\
  \mat{L} &=
            \begin{bmatrix}
              L(\lambda_1) & 0 & \cdots & 0 \\
              0 & L(\lambda_2) & \cdots & 0 \\
              \vdots & \vdots & \ddots & \vdots \\
              0 & 0 & \cdots & L(\lambda_n)
            \end{bmatrix} \Delta \lambda \ .
\end{align*}
There is of course no reason to restrict this setting to a single target object with a single reflectance function $R(\lambda)$. In the general case of $m$ objects, we simply get $\mat{R} \in \RR^{m \times n}$ and $\mat{I}_d \in \RR^{m \times 3}$, i.e., one row per target object. Finally, even more equations can be obtained by taking multiple photographs under different illuminants.

In the context of spectral sensitivity reconstruction, this suggests an experimental setup where the camera under consideration is used to take photographs of objects with known reflectances under known light conditions. Indeed, if $n$ linearly independent equations are obtained this way, then this setup is essentially equivalent to having a monochromator, as we can then just solve Eq.~\eqref{eq:image-model-mat} for $\mat{S}_d$. Unfortunately, this conceptually simple approach suffers from a few practical issues:

\begin{enumerate}
\item The pixel intensities $\mat{I}_d$ in the photographs will always be subject to noise due to the idealized nature of the underlying image formation model in Eq.~\eqref{eq:image-model-cont} that omits the optoelectronics of the camera~\cite{martinez2002calculation,hardeberg1998spectral}, not to mention the discretization error introduced in Eq.~\eqref{eq:image-model-disc}.
\item The illuminant $\mat{L}$ under which the pixel intensities $\mat{I}_d$ are obtained is typically not known; and controlling $\mat{L}$ often requires specialized hardware (see \cref{tab:survey}).
\item Knowing the reflectances $\mat{R}$ requires a special object like a calibrated color chart in the photograph(s).  
\item The linear system does not admit a unique solution as long as $m < n$. Although it is in principle possible to obtain $m = n$ target objects with known linearly independent reflectances, no cheap and convenient options seem to be available on the market today. The common color charts used in photography, for example\textemdash even those containing a large number of different colors\textemdash are printed using just a handful of unique pigments, which essentially limits the possible number of linearly independent reflectances to the number of pigments used.\footnote{The surface on which the colors are printed could contribute with one reflectance as well.}
\end{enumerate}

In our spectral sensitivity reconstruction method, we rely on an alternative, previously overlooked linear system (see \cref{subsec:cmats}) which sidesteps all but the last of these challenges. In order to tame the multitude of candidate solutions arising from the fourth challenge, we employ an autoencoder trained on a database of known spectral sensitivities, which is used to measure how ``realistic'' any given candidate solution is.

\section{Proposed method} \label{sec:proposed-method}

In this section we give a detailed description of our method for indirect spectral sensitivity reconstruction. We begin with a brief introduction to color matrices as well as the linear system of equations that can be obtained form them.

\subsection{Color matrices as a source for equations}
\label{subsec:cmats}

In 2004, Adobe introduced the Digital Negative (DNG), an open-source image file format for storing RAW image information from images created with different cameras~\cite{AdobeDNGSpec}. Each DNG file includes at least one \emph{ColorMatrix} tag, whereas frequently, two matrices per camera can be found. These matrices map camera colors to the CIE Standard Observer, and are calibrated under illuminants with sufficiently far apart correlated color temperatures (CCT), such as illuminant A (tungsten, CCT$\approx$2856K) and D65 (daylight, CCT$\approx$6504K). 

For any device $d$ and illuminant $\mat{L} \in \RR^{n \times n}$, one defines the color matrix $C_d(\mat{L}) \in \RR^{3 \times 3}$ as the least-squares solution to the matrix equation
\[
  \mat{L}\mat{S}_{\sym{XYZ}}C_d(\mat{L}) = \mat{L}\mat{S}_d\ ,
\]
where $\sym{XYZ}$ refers to the CIE XYZ standard observer. Letting $\mat{A}^+ = (\mat{A}^{\top}\mat{A})^{-1}\mat{A}^{\top} \in \RR^{t \times s}$ denote the pseudoinverse of any full-rank matrix $\mat{A} \in \RR^{s \times t}$, we can write
\begin{equation}
  \label{eq:cmats-pseudo-inverse}
  C_d(\mat{L}) = (\mat{L}\mat{S}_{\sym{XYZ}})^+\mat{L}\mat{S}_d \ .
\end{equation}
For multiple illuminants $\mat{L}_1,\dots,\mat{L}_\ell$, we can extend \eqref{eq:cmats-pseudo-inverse} to
\begin{equation}
  \label{eq:cmats-system}
  \begin{bmatrix}
    C_d(\mat{L}_1) \\
    \vdots \\
    C_d(\mat{L}_\ell)
  \end{bmatrix}
  =
  \begin{bmatrix}
    (\mat{L}_1\mat{S}_{\sym{XYZ}})^+\mat{L}_1 \\
    \vdots \\
    (\mat{L}_\ell\mat{S}_{\sym{XYZ}})^+\mat{L}_\ell
  \end{bmatrix} \mat{S}_d \ ,
\end{equation}
which will for us serve as the substitute for \eqref{eq:image-model-mat}. Notice that \eqref{eq:cmats-system} yields $9$ equations for each known calibration matrix, amounting to $9\ell$ equations in total. In practice, calibration matrices are made available only for illuminants A and D65, i.e., $\ell = 2$. It is worth noting that the equations in Eq.~\eqref{eq:cmats-system} are far more practical than those in Eq.~\eqref{eq:image-model-mat}. This is because the former can be constructed entirely from publicly available data, while the latter requires one to obtain photographs of objects with known reflectances under a controlled or estimated illuminant.

\begin{figure*}[ht]
  \centering
  \includegraphics[width=0.95\textwidth]{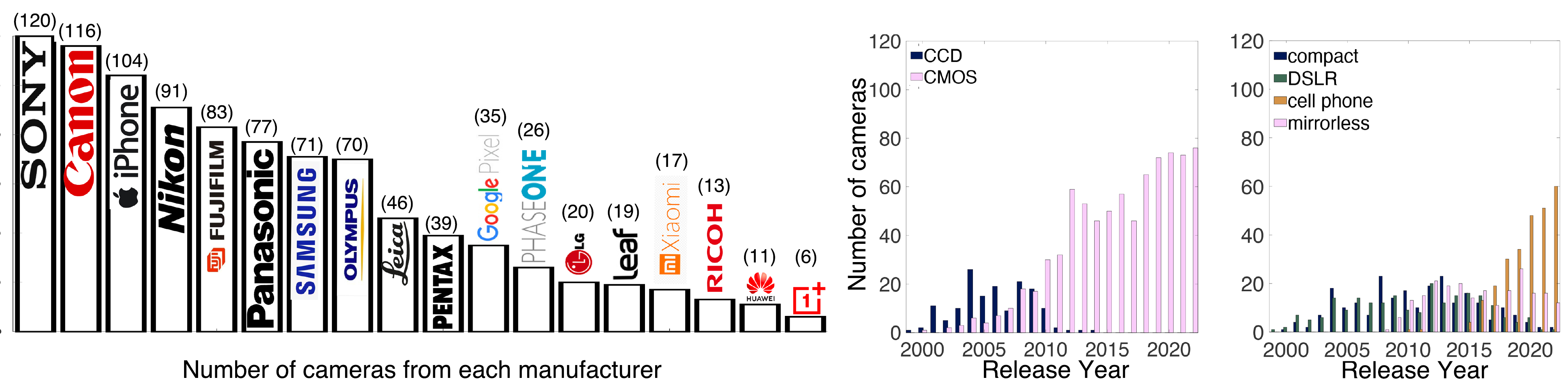}
  \caption{\label{fig:Adobe-summary} \change{We extracted color matrices for $1,013$ digital cameras from the Adobe DNG Converter and manually compiled metadata for each camera including camera type (e.g., DSLR, compact, mirrorless, cell phone), sensor type (CMOS or CCD), sensor size (in mm), and release year. Manufacturers with fewer than 5 cameras were omitted. See Supplementary Materials for details.}}
  \end{figure*}

Adobe matrices can be acquired as follows: For a given RAW image, the available color calibration matrices will be appended to the image metadata when the image is processed through the Adobe DNG Converter. These matrices can then be obtained via a viewer, e.g., \emph{ExifTool}~\cite{exiftool}. This option still does not necessitate having physical access to the camera in question, as example RAW images from many cameras can be freely found online (see~\cite{RawSamples} for an excellent source). We obtained the color matrices for $1,013$ cameras for which the Adobe DNG Converter version 12.4.0.555 supports, using the code provided by~\cite{AdobeMatrices}. \change{Some statistics for these cameras are shown in~\cref{fig:Adobe-summary} for manufacturers who had $\geq 5$ cameras in the dataset.}

\subsection{Optimization problem}
\label{subsec:optimization}


To begin with, assume that we describe the sought spectral sensitivities $\mat{S}_d \in \RR^{n \times 3}$ as a solution to a linear system $\mat{A}\mat{S}_d = \mat{B}$ for some known matrices $\mat{A} \in \RR^{m \times n}$ and $\mat{B} \in \RR^{m \times 3}$. For example, this linear system could be \eqref{eq:image-model-mat}, or \eqref{eq:cmats-system}, or a combination of the two. Independently of which linear system it is, we can assume that is underdetermined, i.e., $m < n$; otherwise, we can just solve the system and there is nothing more to do. To address the insufficient number of equations at our disposal, we follow the usual practice of inventing a function $F$ that maps any potential spectral sensitivity candidate $\mat{S} \in \RR^{n \times 3}$ to a single real number that is supposed to indicate how ``realistic'' this candidate is. For example, if we assume that spectral sensitivities are generally rather smooth, then we could choose $F(\mat{S}) = \norm{\mat{D}\mat{S}}_2$, where $\mat{D} \in \RR^{n \times n}$ is the second derivative matrix and $\norm{\cdot}_2$ is the Frobenius norm; variants of this were explored in e.g., \cite{barnard2002camera} and \cite{huynh2014recovery}.

Being equipped with the aforementioned system of equations as well as the function $F$, we combine the two in in the formulation of the final optimization problem
\begin{equation}
  \label{eq:optimization-framework}
  \argmin_{\mat{S} \in \Omega} \underbrace{D(\mat{A}\mat{S}, \mat{B})}_{\text{specific term}}
  \hspace{5pt} + \hspace{-8pt} \underbrace{F(\mat{S})}_{\text{universal term}} \hspace{-10pt} ,
\end{equation}
where $\Omega \subseteq \RR^{n \times 3}$ is the feasible domain and $D$ is some distance function on $\RR^{m \times 3}$. In a sense, the whole field of indirect spectral sensitivity reconstruction can be viewed as the search for the right combination of the two terms in Eq.~\eqref{eq:optimization-framework}, hoping that an optimization algorithm would converge to a solution not too far off from the sought $\mat{S}_d$. Before proceeding to our take on this framework, we ought to mention that the system of equations in the specific term does not have to be linear; an example of a quadratic system can be found in \cite{finlayson1998recovering}. Furthermore, the two terms can of course be summed in a weighted manner\textemdash we have omitted the weight coefficients for a cleaner presentation.

Having presented the general optimization framework for spectral sensitivity reconstruction, let us introduce our proposed realization of it. For the specific term, we rely on the linear system in Eq.~\eqref{eq:cmats-system} based on color matrices, choosing
\[
  \mat{A}
  = \change{\begin{bmatrix}
    \mat{A}_1 \\
    \mat{A}_2
  \end{bmatrix}}
  =
  \begin{bmatrix}
    (\mat{L}_{\sym{A}}\mat{S}_{\sym{XYZ}})^+\mat{L}_{\sym{A}} \\
    (\mat{L}_{\sym{D65}}\mat{S}_{\sym{XYZ}})^+\mat{L}_{\sym{D65}}
  \end{bmatrix} , \
  \mat{B}
  = \change{\begin{bmatrix}
    \mat{B}_1 \\
    \mat{B}_2
  \end{bmatrix}}
  =
  \begin{bmatrix}
    C_d(\mat{L}_{\sym{A}}) \\
    C_d(\mat{L}_{\sym{D65}})
  \end{bmatrix} \ ,
\]
where $\mat{L}_{\sym{A}}$ and $\mat{L}_{\sym{D65}}$ denote the illuminants A and D65 respectively (see \cref{subsec:cmats}). For the feasible domain and the distance function we respectively use
\begin{align*}
  &\Omega = \{\mat{S} \mid \mat{S} \ne \mat{0} \text{ has no negative entries}\} \quad \text{and} \\
  &\change{D(\mat{A}\mat{S}, \mat{B}) = \sum_{i=1}^2\angle(\mat{A}_i\mat{S}, \mat{B}_i) ,}
    \quad \text{where} \\
  &\angle(\mat{U}, \mat{V}) \triangleq \arccos \big( \frac{\mat{U} \cdot \mat{V}}{\norm{\mat{U}}_2\norm{\mat{V}}_2} \big) \ ,
\end{align*}
and $\mat{U} \cdot \mat{V}$ denotes the dot product of $\mat{U}$ and $\mat{V}$ seen as vectors. Finally, our universal term takes the form
\begin{equation}
  \label{eq:our-specific-term}
  F(\mat{S}) = \angle(\mat{S}, A_{\vec{w}}(\mat{S})) \ ,
\end{equation}
where $A_{\vec{w}}$ is an autoencoder with weights $\vec{w}$ trained on a database $\mathcal{S} \subset \RR^{n \times 3}$ of known spectral sensitivities. Details about the chosen architecture of $A$, as well as the training process, can be found in \cref{subsec:autoencoder}.

\subsection{Autoencoder}
\label{subsec:autoencoder}


Letting $\mathcal{S} \subset \RR^{n \times 3}$ denote a database of ground-truth spectral sensitivities, the weights $\vec{w}$ of our outoencoder $A_{\vec{w}}$ are obtained as a solution to the optimization problem
\begin{equation}
  \label{eq:autoencoder-training}
  \argmin_{\vec{w}} \frac{1}{|\mathcal{S}|}\sum_{\mat{S} \in \mathcal{S}} \Delta(\mat{S}, A_{\vec{w}}(\mat{S})) \ ,
\end{equation}
where for any $\mat{U} = [\mat{U}_{\sym{r}}, \mat{U}_{\sym{g}}, \mat{U}_{\sym{b}}]$ and $\mat{V}= [\mat{V}_{\sym{r}}, \mat{V}_{\sym{g}}, \mat{V}_{\sym{b}}]$ in $\RR^{n \times 3}$
\[
  \Delta(\mat{U},\mat{V}) = \left\lVert \left[ \frac{\mat{U}_k - \mat{V}_k}{\norm{\mat{U}_k}_2} \right]_{k=\sym{r},\sym{g},\sym{b}} \right\rVert_2 \ .
\]
The choice for $\Delta$ might come across as somewhat peculiar, but it simply turns out to work better than $\angle(\cdot, \cdot)$, which would be more consistent with our choice of the universal term in Eq.~\eqref{eq:our-specific-term}. Another choice could be $\Delta(\mat{U}, \mat{V}) = \norm{\mat{U} - \mat{V}}_2$, which is essentially what we are doing, except that the scaling by $\norm{\mat{U}_k}_2^{-1}$ is introduced to make $A_{\vec{w}}$ assign roughly equal importance to the three color channels.

The description thus far constitutes the conceptual core of of our training process, but we have omitted two important details, both related to the small size of our database $\mathcal{S}$ (see \cref{tab:cameras}). First and foremost, it has shown to be absolutely crucial for our method to use regularization in order to avoid dramatic overfitting; in particular, we use weight decay and dropout (for parameter choices see \cref{tab:autoencoder-training}). Secondly, the stability of our method seems to improve with addition of modest data augmentation. During training, every time we draw a ground-truth sensitivity $\mat{S}$ from $\mathcal{S}$, we randomly scale and ``roll'' its channels by applying the map $\mat{S} \mapsto \mat{G}\mat{S}\mat{H}$ where $\mat{H} \in \RR^{3 \times 3}$ is a diagonal matrix whose diagonal entries are drawn uniformly and independently from the interval $[h, 1]$ for a real-valued parameter $0 < h \le 1$. The matrix $\mat{G} \in \RR^{n \times n}$ is a special kind of random permutation matrix whose $i$-th column is $0$ in every entry except for the index
\[
  (i - 1 + u \mod n) + 1 \ ,
\]
where the value is $1$; here $u$ is a random integer drawn uniformly and independently for every column of $\mat{G}$ from the set $\{-g,\dots,g\}$, where $0 \le g < n$ is an integer-valued parameter.  The two parameters $h$ and $g$ are chosen by the user and decide the level of data augmentation. Intuitively, $h$ decides by how much we can randomly scale the channel peaks up and down, while $g$ decides by how much we can shift them left and right. Our choice for these parameters can be found in \cref{tab:autoencoder-training}, while the neural network architecture of the autoencoder $A_{\vec{w}}$ is presented in \cref{tab:autoencoder-architecture}. \change{Finally, we ought to mention that all ground-truth sensitivities $\mat{S} \in \mathcal{S}$ were normalized by their maximal value (before augmentation).}

\begin{table}[h]
  \caption{\label{tab:autoencoder-architecture} Architecture of the autoencoder $A_{\vec{w}}$. Information travels from top to bottom. In the table $N = 3n$, and $p$ indicates that dropout with the specified retention probability was applied after the corresponding layer during training.}\vspace{-12pt}
\begin{center}
\begin{tabular}{lccl@{\hskip -10pt}l}
  \hline
  {\bf Layer type} & {\bf Inputs} & {\bf Outputs} & {\bf Activation} & {}\\
  \hline
  Linear $(p=0.2)$ & $N$ & $4N$ & ReLU & \multirow{3}{*}{$\left.\begin{array}{l}
                                                        \\ \\                                                        
                                                      \end{array}\right\rbrace\text{Encoder}$} \\
  Linear $(p=0.5)$ & $4N$ & $2N$ & ReLU & \\
  Linear & $2N$ & $6$ & None & \\
  \hline
  Linear & $6$ & $2N$ & ReLU & \multirow{3}{*}{$\left.\begin{array}{l}
                                                        \\ \\                                                        
                                                      \end{array}\right\rbrace\text{Decoder }$} \\
  Linear $(p=0.5)$ & $2N$ & $4N$ & ReLU & \\
  Linear & $4N$ & $N$ & None & \\
  \hline
\end{tabular}
\end{center}
\end{table}

\begin{table}[h]
  \caption{\label{tab:autoencoder-training} Training details for weights $\vec{w}$ in \eqref{eq:autoencoder-training}.  Explanations for channel scaling and rolling (data augmentation) can be found in \cref{subsec:autoencoder}.}
\begin{center}
\begin{tabular}{ll}
  \hline
  {\bf Parameter} & {\bf Value} \\
  \hline
  Framework & {\sf PyTorch v1.11.0} \\
  Optimizer & {\sf SGD} \\
  Channel scaling $(h)$ & $2 \cdot 10^{-1}$ \\
  Channel rolling $(g)$ & $2$ \\
  Learning rate $(lr)$ & $10^{-1}$ \\
  Momentum & $5 \cdot10^{-1}$ \\
  Weight decay & $10^{-4}$ \\
  Scheduler & {\sf ReduceLROnPlateau} \\
  Scheduler decay & $5 \cdot10^{-1}$ \\
  Scheduler patience & $2 \cdot 10^3$\\
  Stop criterion & $lr < 10^{-5}$ \\
  \hline
\end{tabular}
\end{center}
\end{table}

\subsection{Ground-truth database}
\label{subsec:gt-database}

  The database $\mathcal{S}$ consists of 51 spectral sensitivity functions that we compiled from the literature, which were reported to have been obtained directly using monochromatic light (\cref{tab:cameras}). If the authors did not make their data available in a table, we digitized them from plots using DataThief~\cite{datathief}.

  We excluded some cameras from the sources listed. Specifically, we excluded all cameras for which Adobe DNG Converter version 12.4.0.555 did not provide a \emph{ColorMatrix} tag. In some cases, the presented curves were scaled by the authors so as to make the maximal value in each channel equal to one\textemdash these were excluded from our dataset due to the across-channel nature of our approach. If a camera's curves showed sensitivity to a wider range than the VIS (i.e., UV or IR filters were removed), we omitted that camera. Finally, we also excluded discontinued cameras with atypical curves, such as the Kodak DCS series (a poignant outcome given their pioneering role in the history of digital photography).

\begin{figure}[ht]
  \centering
  \includegraphics[width=0.5\textwidth]{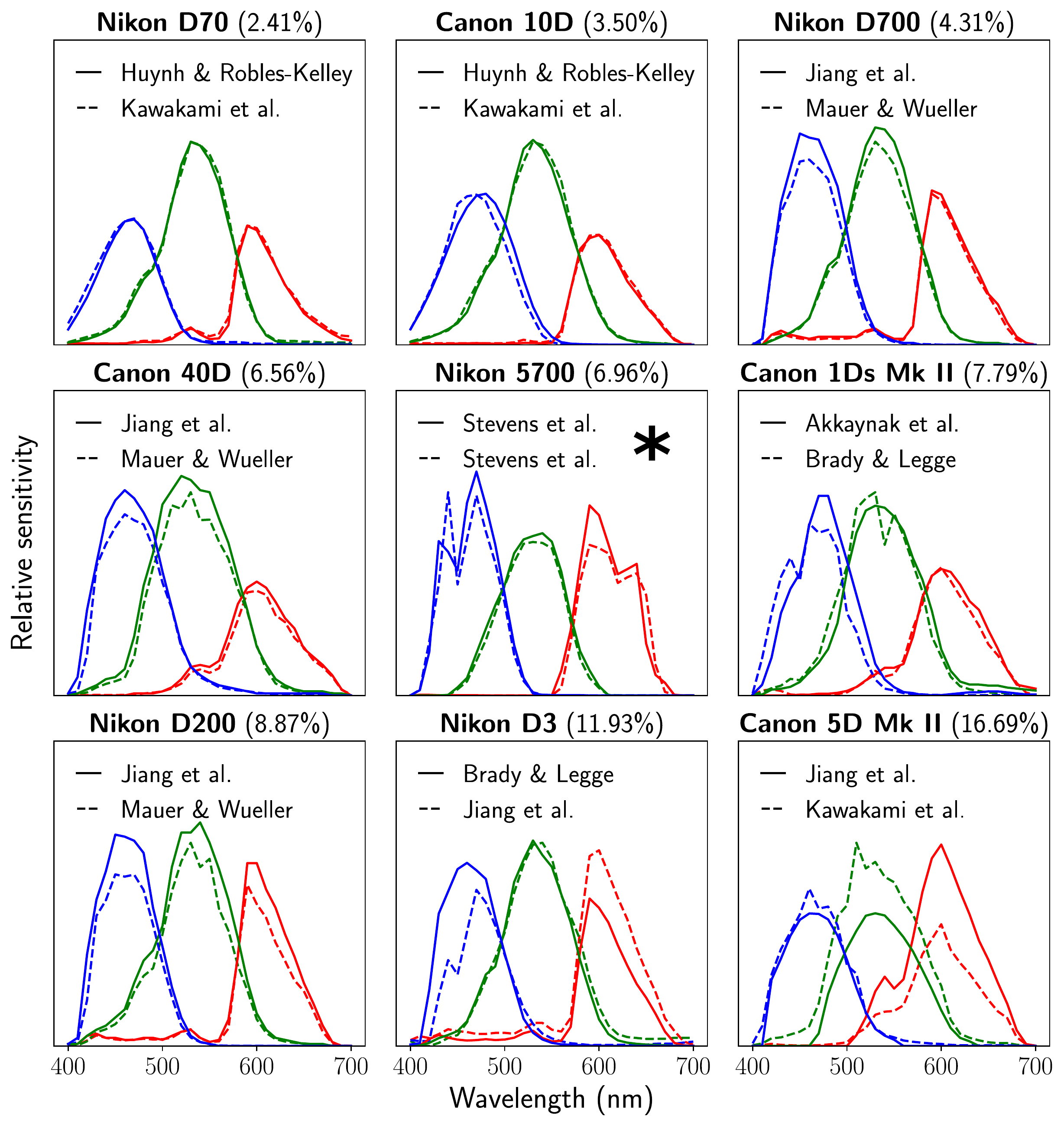}
  \caption{\label{fig:gt-curves-8} Cameras in our database for which two ground-truth derivations exist. Percentages indicate relative full-scale error between estimations reported in two different publications for each camera. These differences probably arise from measurement errors and uncertainties associated with the experimental setup employed by each party. Note, however, that the case for the two Nikon 5700 Coolpix cameras (*) is slightly different. These cameras were reported to be purchased at the same time and their spectral sensitivity measurements were made with the same experimental setup~\cite{stevens2007using}. Thus, potentially, the $6.96\%$ error between their spectral sensitivities primarily represents the error resulting from the imperfections of the manufacturing process and/or the variability of the sensor materials.}
  \end{figure}

  Among the resulting set of 51, we had 9 cameras whose sensitivities were derived by more than one source (\cref{fig:gt-curves-8}). The relative full-scale errors between these duplicates range between $\approx2.4-17\%$,  which are likely due to measurement errors and uncertainties; this can be quite large for some wavelengths, as demonstrated by Darrodi et al.~\cite{darrodi2015reference}. It is also possible, however, that two copies of the same make and model camera have quantifiably different spectral sensitivities due to the imperfections of the manufacturing process. This indeed seems to be the case for two Nikon Coolpix 5700 cameras that were reportedly purchased simultaneously and characterized using the same experimental setup by Stevens et al.~\cite{stevens2007using} (note that Adobe does not provide matrices for this camera but we retained it in our database). 

\begin{table}[h]
    \caption{Our compiled ground-truth spectral sensitivity database of 51 cameras. All original sources report estimation using a monochromator, except for those that used a liquid crystal modulator$^+$, diffraction grating$^*$, and narrowband filters$^\dag$ to obtain monochromatic light.}\label{tab:cameras}\vspace{-12pt}
\begin{center}
\begin{tabular}{p{1.7in}|c}
\hline
\textbf{Camera make \& model} & \textbf{Source} \\
\hline
Canon 1Ds II & \multirow{1}{*}{Akkaynak et al.~\cite{akkaynak2011using}} \\ \hline
Panasonic DMCL X5,  Sony NEX7 & \multirow{2}{*}{Berra et al.~\cite{berra2015estimation}}  \\ \hline 
Canon 1Ds II,  Nikon D3 & \multirow{1}{*}{Brady \& Legge~\cite{brady2009camera}} \\ \hline
Samsung Galaxy S8 & \multirow{1}{*}{Burggraaff et al.~\cite{burggraaff2019standardized}} \\ \hline
Canon 10D, Nikon D70 & \multirow{1}{*}{Huynh \& Robles-Kelley~\cite{huynh2007comparative}$^+$} \\ \hline
Canon 1D III, Canon 20D, Canon 300D, Canon 40D, Canon 500D, Canon 50D, Canon 5D II, Canon 600D, Canon 60D, Nikon D200, Nikon D3, Nikon D300s, Nikon D3X, Nikon D40, Nikon D50, Nikon D5100, Nikon D700, Nikon D80, Nikon D90, Olympus EPL2, Pentax K5, Pentax Q, Sony NEX5N & \multirow{10}{*}{Jiang et al.~\cite{jiang2013space}}\\ \hline
Canon 10D, Canon 5D II, Canon 5D,  Nikon D1x, Nikon D70& \multirow{2}{*}{Kawakami et al.~\cite{kawakami2013camera}} \\ \hline
Canon 400D & \multirow{1}{*}{Lebourgeois et al.~\cite{lebourgeois2008can}$^*$} \\ \hline
Canon 40D,  Leica M8, Nikon D200, Nikon D700, Panasonic DMCL-X3 & \multirow{3}{*}{Mauer \&Wueller \cite{mauer2009measuring}$^\dag$} \\ \hline
Nikon 5700 (2 copies) & \multirow{1}{*}{Stevens et al.~\cite{stevens2007using}} \\ \hline
iPhone	8, iPhone	X, iPhone	11, iPhone	12 Pro Max, Samsung	Galaxy S7 Edge, Samsung	Galaxy S9, Samsung	Galaxy S20 & \multirow{5}{*}{Tominaga et al.~\cite{tominaga2021measurement}} \\ \hline
\end{tabular}
\end{center}
\end{table}

\section{Results}
\label{sec:results}

When it comes to indirect methods for spectral sensitivity reconstruction, it is generally difficult\textemdash if not impossible\textemdash to formulate any concrete theoretical guarantees about reconstruction accuracy. The main reason for this is that all contemporary models for the space of spectral sensitivity functions substitute the lacking theoretical understanding of this space with convenient assumptions that may or may not hold in practice. In the case of our method, it is assumed that this space can be learned by an autoencoder with a particular neural network architecture, which doesn't make theoretical analysis any easier. In acknowledgment of this, we proceed with the standard approach of assessing the effectiveness of our method by empirical validation.

\begin{figure}[ht]
  \centering
  \includegraphics[width=0.5\textwidth]{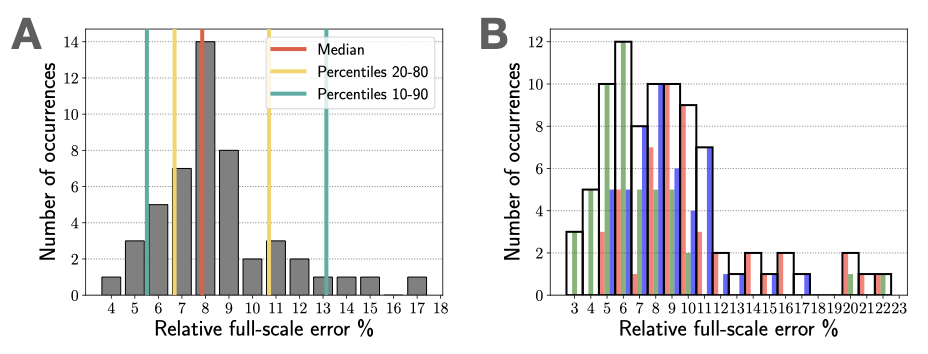} 
  \caption{\label{fig:loov-hist} \textbf{A)} Histogram of relative full-scale errors observed during leave-one-out validation of our method. The two ground-truths of Nikon 5700 from Stevens et al.~\cite{stevens2007using} were excluded from the predictions as well as from training data due to missing color matrices. \change{\textbf{B)} Breakdown of errors per color channel\textemdash the median errors are $9.3\%$, $6.3\%$ and $8.3\%$ for red, green and blue channels, respectively.}}
\end{figure}

\begin{figure}[h]
  \centering
    \includegraphics[width=0.5\textwidth]{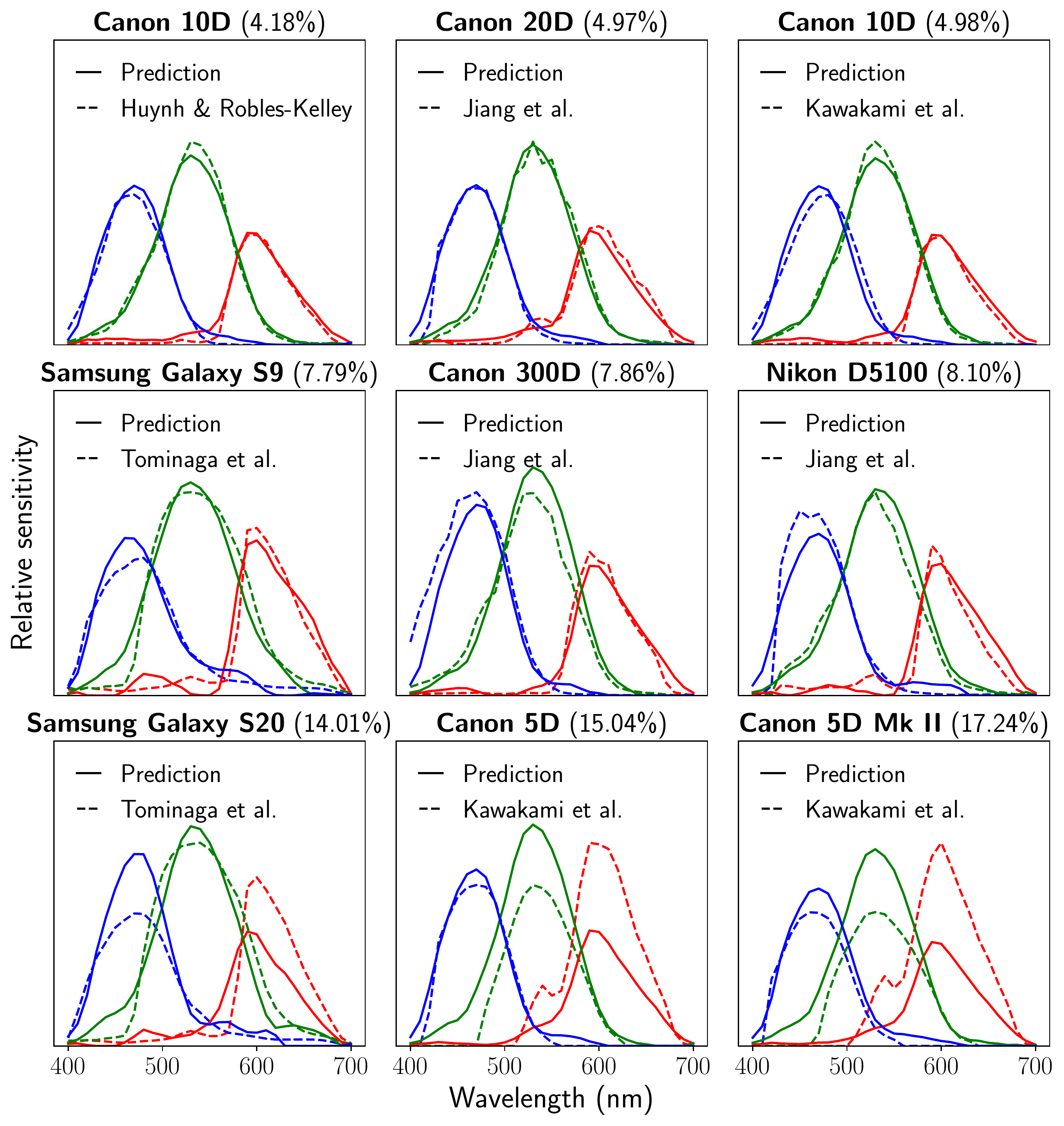}
    \caption{\label{fig:loov-samples} Our best (first row), median (middle row), and worst (bottom row) results for the 51 cameras in our ground-truth set. The rest of the results for the ground-truth dataset, as well as predictions for $1,000+$ cameras are given in Supplementary Material.}
  \end{figure}

Our validation procedure is a form of leave-one-out validation and can be summarised as follows: For every device $d$ represented in our database $\mathcal{S} \subset \RR^{n \times 3}$, we train an autoencoder $A^{(d)}$ on $\mathcal{S} \setminus \{\mat{S}_d\}$. We then proceed by solving the optimization problem discussed in \cref{subsec:optimization}, i.e.,
\begin{align*}
  &\argmin_{\mat{S}} \hspace{5pt} \alpha \change{\sum_{i=1}^2\angle(\mat{A}_i\mat{S}, \mat{B}_i)}
    + \beta \angle(\mat{S}, A^{(d)}(\mat{S})) , \text{ where} \\
  &\change{\begin{bmatrix}
       \mat{A}_1 \\
       \mat{A}_2
     \end{bmatrix}}
    =
    \begin{bmatrix}
      (\mat{L}_{\sym{A}}\mat{S}_{\sym{XYZ}})^+\mat{L}_{\sym{A}} \\
      (\mat{L}_{\sym{D65}}\mat{S}_{\sym{XYZ}})^+\mat{L}_{\sym{D65}}
    \end{bmatrix} \quad \text{and} \quad
    \change{\begin{bmatrix}
       \mat{B}_1 \\
       \mat{B}_2
      \end{bmatrix}}
    = \begin{bmatrix}
      C_d(\mat{L}_{\sym{A}}) \\
        C_d(\mat{L}_{\sym{D65}})
    \end{bmatrix} \ ,
\end{align*}
and comparing the found solution $\widehat{\mat{S}}_d$ to the ground-truth $\mat{S}_d$. Here,  $\alpha, \beta \ge 0$ are just weight coefficients; their values, as well as other optimization parameters can be found in \cref{tab:optimization-params}. The comparison is done similarly to \cite{zhu2020spectral} with respect to the average \emph{relative full-scale error} across the three color channels, i.e.,
\begin{align*}
  &\text{RE}(\widehat{\mat{S}}_d, \mat{S}_d) = \frac{1}{3}\sum_{k \in \{\sym{r}, \sym{g}, \sym{b}\}}
  \frac{\text{RMSE}(\widehat{\mat{S}}_d^{[*,k]}, \mat{S}_d^{[*,k]})}{\max \mat{S}_d^{[*,k]}} \ , \quad \text{where} \\
  &\text{RMSE}(\widehat{\mat{S}}_d^{[*,k]}, \mat{S}_d^{[*,k]}) = \sqrt{\frac{1}{n}\sum_{i = 1}^n (\widehat{\mat{S}}_d^{[i,k]} - \mat{S}_d^{[i,k]})^2}
\end{align*}
is the root mean square error; here the superscript $[i,k]$ means taking the entry in the $i$-th row and the $k$-th column, while $[*,k]$ refers to the $k$-th column in its entirety.  As mentioned, our database $\mathcal{S}$ contains a few duplicate ``ground-truth'' sensitivities for the same camera. Whenever this is the case, we leave out \emph{all} of these duplicates when training the autoencoder during the validation (but errors are still included). The resulting error histogram can be found in \cref{fig:loov-hist}\, and our best, median, and worst results are shown in ~\cref{fig:loov-samples}. \change{We also performed validation where we left out one brand at a time\textemdash the results were similar to those in \cref{fig:loov-hist} (e.g. $7.08\%$ median error across all Canon cameras).} The rest of the results are in the Supplementary Material.

 It is difficult to compare this error range to other methods, as there is no consensus in the literature about which error metric should be used, and many of the common error metrics are scale-dependent. One value with which we can compare is the reported $8.54 \%$ average relative full-scale error across channels from Zhu et al.~\cite{zhu2020spectral} for Canon EOS 600D; this is slightly worse than our error of $6.87 \%$ with respect to the ground-truth from Jiang et al.~\cite{jiang2013space}. The method of Zhu et al.~\cite{zhu2020spectral} required a camera and specialized hardware (\cref{tab:survey}).

  We wondered if reconstruction accuracy would improve if we also used a color chart photo for a given camera. We used the outdoor images from the NUS dataset~\cite{cheng2014illuminant} all of which contained a ColorChecker, and included the equations from the image formation model \eqref{eq:image-model-mat} into the specific term of our optimization problem. We approximated the light by the CIE standard daylight model, parameterized by CCT, which we treated as an additional unknown to be solved for in our optimization problem. Our reconstruction accuracy did not improve by any significant amount ($< 0.5 \%$). Most likely, this was because the color chart offered little new information that was not captured by the color matrices.


  \begin{table}[h]
  \caption{\label{tab:optimization-params} Optimization parameters used to produce \cref{fig:loov-hist}.}
\begin{center}
\begin{tabular}{ll}
  \hline
  {\bf Parameter} & {\bf Value} \\
  \hline
  Framework & {\sf PyTorch v1.11.0} \\
  Optimizer & {\sf SGD} \\
  Specific term coeff. $(\alpha)$ & $10^2$ \\
  Universal term coeff. $(\beta)$ & $2 \cdot 10^{-1}$ \\
  Learning rate $(lr)$ & $8 \cdot 10^{-4}$ \\
  Scheduler & {\sf ReduceLROnPlateau} \\
  Scheduler decay & $5 \cdot 10^{-1}$ \\
  Scheduler patience & $2 \cdot 10^3$ \\
  Stop criterion & $lr < 4 \cdot 10^{-4}$ \\
  \hline
\end{tabular}
\end{center}
\end{table}

\section{Discussion}
We introduced a zero-cost framework with which to reasonably estimate the spectral sensitivities of consumer cameras. Our reconstruction errors ranged from $4-17\%$ with a median of $\approx8\%$ (\change{\cref{fig:loov-hist}}), which is comparable to the errors \change{that might arise due to sensor fabrication flaws or irregularities when two copies of the same are being manufactured (see Nikon 7500 data in \cref{fig:gt-curves-8}}). Since our framework relies on publicly available information, our current results can be improved as more, and higher-quality data become available, \change{for instance, when more ground-truth curves are published by researchers, or color matrices for more illuminants are derived by Adobe or others}.  We also provided predictions of camera responses for $1,000+$ cameras \change{for the use of the scientific community.}

A lot of value can be added to scientific research with the knowledge of the spectral sensitivities of the camera at hand. This information can enable objectivity and repeatability of color capture, better colorimetric mapping, and create the possibility of linking \texttt{RGB} values to spectral quantities in the physical world. \change{We demonstrated some of these cases in ~\cref{fig:fig1} where we modified the implementations of previously published methods to accommodate the spectral sensitivities of the cameras used. Moreover, having the spectral sensitivities reduces the number of unknowns in a given problem by $3n$ (where $n$ represents the number of chosen wavelength steps), thereby simplifying solutions in numerous scenarios.}

\change{For example,} \textbf{illuminant spectrum estimation} can become a simpler problem with a more accurate solution that recovers the \emph{spectrum} of the physical illuminant, not just the camera-specific \texttt{RGB} white point \change{(\cref{fig:fig1}\textbf{A}).} \change{Additionally, as suggested by Cheng et al.~\cite{cheng2014illuminant}}, algorithms that use specular and shadow pixels to estimate the illuminant through distinguishing between specularities and bright surfaces, and shades and dull surfaces, can be refined. 

Similarly, \textbf{reflectance spectrum estimation} also becomes simpler, especially outdoors where light can be represented using the one-parameter CIE daylight model. Underwater, image formation is complex, but in the cases where haze can be neglected and~\cref{eq:image-model-cont} applies, the spectrum of the \textbf{diffuse downwelling attenuation coefficient} can be recovered using a color chart and the CIE daylight model (\change{\cref{fig:fig1}\textbf{B})}. It should be kept in mind, however, that metamerism and the low-dimensionality of spectral reflectances on most color charts still make the estimation of spectral quantities challenging \change{(see Supplementary Material)}.

Having access to the spectral sensitivities of \emph{many} cameras, not all of which may be at hand, can also add a lot value to science. For example, \change{realistic} training and/or augmentation data can be generated for \textbf{image colorization}, \textbf{jpg-to-raw}~\cite{conde2022reversed}, and \textbf{raw-to-raw}~\cite{nguyen2014raw} mapping problems. \textbf{Spectral super-resolution}, a task which obtains hyper-spectral images from a single RGB image and \change{generally requires camera curves to be known~\cite{nguyen2014training}}, will be possible to apply to any consumer camera image. Similarly, many cameras will suddenly become candidates for \textbf{hyper-spectral super-resolution}, the task of fusing a hyperspectral image with low spatial resolution with an RGB image of high spatial resolution, \change{which also requires camera responses to be known~\cite{akhtar2014sparse,akhtar2016hierarchical,kawakami2013camera}}. It will also be trivial follow the method of Oh et al.~\cite{oh2016yourself} to build a Do-it-Yourself hyperspectral imaging system by combining several different consumer cameras, \change{since the dataset we release contains the peak wavelengths of (almost) all the cameras in today's market.}

Perhaps the most curious application of our predicted camera responses is \textbf{time travel rephotography}, which is the projection of  faded, black-and-white photos into the space of modern-day high resolution sensors~\cite{luo2021time}. It will be possible to re-make iconic photographs, like that of Abraham Lincoln, in the color space of any modern-day digital camera, simulating that camera's in-camera photofinishing workflow using the pipeline from~\cite{karaimer2016software}, and even comparing how each camera in the market would have portrayed the notable persona.

\begin{figure}[h]
  \centering
    \includegraphics[width=0.45\textwidth]{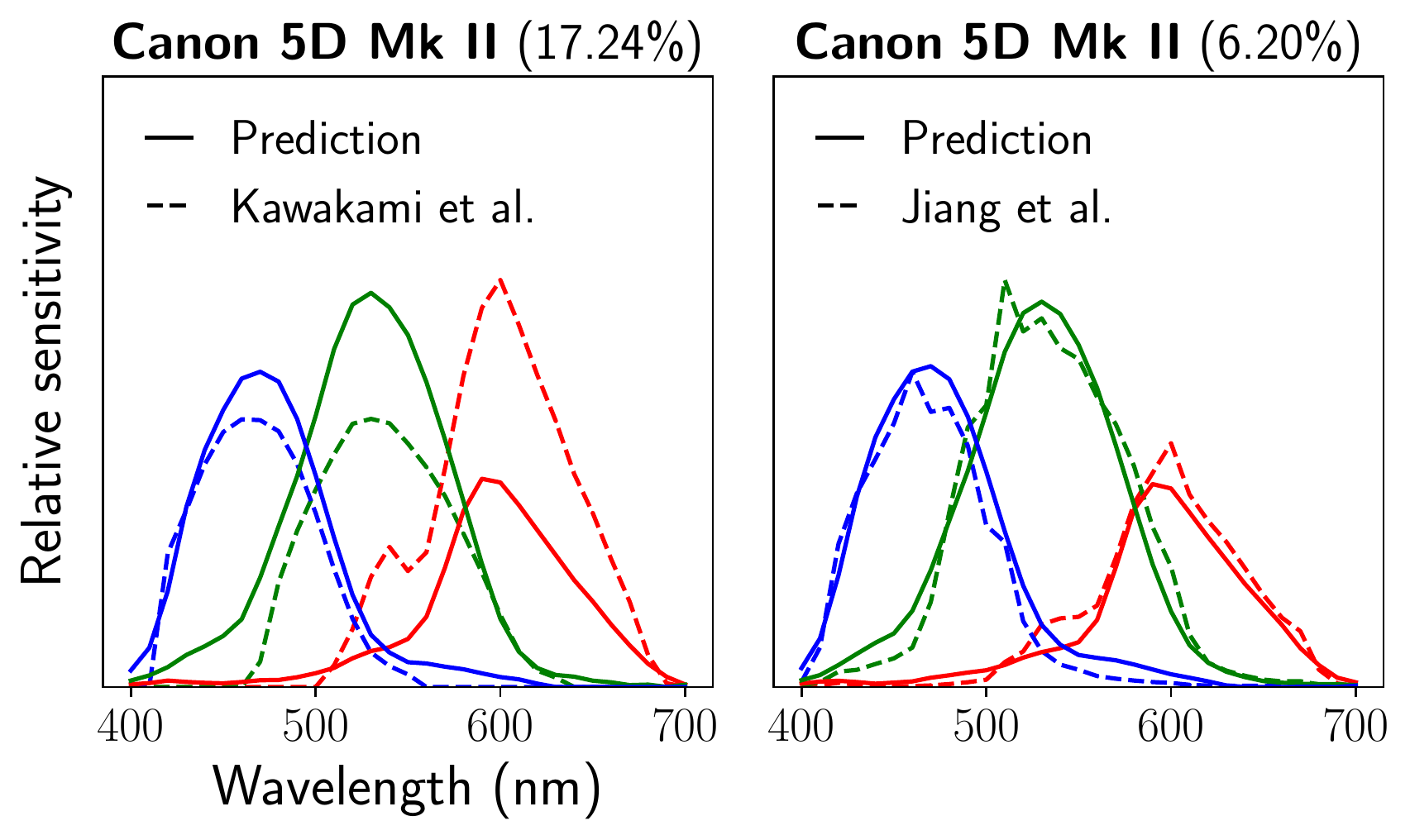}
    \caption{\label{fig:kawakami-busted} \change{Leave-one-out validation results for two different ``ground-truths'' of Canon 5D Mk II. The large discrepancy in reconstruction error could suggest and experimental error in Kawakami et al., but further investigation is needed for a definitive conclusion.}}
  \end{figure}

While we are excited about all the capabilities our work will enable, we acknowledge that our method has limitations. First of all, we noticed large differences in the reconstruction errors on the same camera due to the \emph{source} of derivation (\cref{fig:gt-curves-8}), which means that ``ground-truth" information should be interpreted with caution. For example, our reconstruction for Canon 5D Mk II yielded only $6.20\%$ error relative to the ground-truth from Jiang et al.~\cite{jiang2013space}, but $17.24\%$ error relative to ground-truth from Kawakami et al.~\cite{kawakami2013camera}, which is actually our worst result (see \cref{fig:kawakami-busted}).
Does the sensor in this Canon model show such a large variation? Or was Kawakami's derivation flawed? We cannot answer, and we caution the reader to be aware of potential measurement errors and uncertainties when using publicly available data. Also, we assume, as does the Adobe DNG Converter, that every copy of a given make and model camera has identical spectral sensitivities. The case of Canon 5D Mk II could be one example where this assumption fails. However, we believe that in practice differences between sensors of identical cameras are small, and that the Nikon 5700 example in~\cref{fig:gt-curves-8} is more typical of this situation, so our general approach is justified.

Lastly, we did not have ground-truth data for most of the $1,000+$ cameras we presented, because if such data were available, we would not need to write this paper. Despite the lack of ground-truth, we have confidence in the peak wavelengths and the general shapes of the reconstructed curves due to the way Adobe's color matrices are derived. Even without ground-truth, this information can already be useful, serving as a feasible initial guess for problems where camera responses are unknown, or potentially guiding the purchase of a camera if sensitivity for certain wavelengths are desired.

\section*{Acknowledgments}
This research was funded by Israel Science Foundation grant $\#1055/22$ and the Schmidt Marine Technology Partners. We thank Olena Kovalenko for assistance with data analysis and visualization, and four anonymous reviewers for their helpful comments. 

\addcontentsline{toc}{section}{Acknowledgment} 

\bibliographystyle{IEEEtran}
\bibliography{references}

\begin{thebibliography}{10}
\providecommand{\url}[1]{#1}
\csname url@samestyle\endcsname
\providecommand{\newblock}{\relax}
\providecommand{\bibinfo}[2]{#2}
\providecommand{\BIBentrySTDinterwordspacing}{\spaceskip=0pt\relax}
\providecommand{\BIBentryALTinterwordstretchfactor}{4}
\providecommand{\BIBentryALTinterwordspacing}{\spaceskip=\fontdimen2\font plus
\BIBentryALTinterwordstretchfactor\fontdimen3\font minus
  \fontdimen4\font\relax}
\providecommand{\BIBforeignlanguage}[2]{{%
\expandafter\ifx\csname l@#1\endcsname\relax
\typeout{** WARNING: IEEEtran.bst: No hyphenation pattern has been}%
\typeout{** loaded for the language `#1'. Using the pattern for}%
\typeout{** the default language instead.}%
\else
\language=\csname l@#1\endcsname
\fi
#2}}
\providecommand{\BIBdecl}{\relax}
\BIBdecl

\bibitem{burggraaff2019standardized}
O.~Burggraaff, N.~Schmidt, J.~Zamorano, K.~Pauly, S.~Pascual, C.~Tapia,
  E.~Spyrakos, and F.~Snik, ``Standardized spectral and radiometric calibration
  of consumer cameras,'' \emph{Optics express}, vol.~27, no.~14, pp.
  19\,075--19\,101, 2019.

\bibitem{akkaynak2014use}
D.~Akkaynak, T.~Treibitz, B.~Xiao, U.~A. G{\"u}rkan, J.~J. Allen, U.~Demirci,
  and R.~T. Hanlon, ``Use of commercial off-the-shelf digital cameras for
  scientific data acquisition and scene-specific color calibration,''
  \emph{JOSA A}, vol.~31, no.~2, pp. 312--321, 2014.

\bibitem{kim2012new}
S.~J. Kim, H.~T. Lin, Z.~Lu, S.~S{\"u}sstrunk, S.~Lin, and M.~S. Brown, ``A new
  in-camera imaging model for color computer vision and its application,''
  \emph{IEEE Transactions on Pattern Analysis and Machine Intelligence},
  vol.~34, no.~12, pp. 2289--2302, 2012.

\bibitem{paris2009fast}
S.~Paris and F.~Durand, ``A fast approximation of the bilateral filter using a
  signal processing approach,'' \emph{International journal of computer
  vision}, vol.~81, pp. 24--52, 2009.

\bibitem{guo2020zero}
C.~Guo, C.~Li, J.~Guo, C.~C. Loy, J.~Hou, S.~Kwong, and R.~Cong,
  ``Zero-reference deep curve estimation for low-light image enhancement,'' in
  \emph{Proceedings of the IEEE/CVF conference on computer vision and pattern
  recognition}, 2020, pp. 1780--1789.

\bibitem{zhu2017unpaired}
J.-Y. Zhu, T.~Park, P.~Isola, and A.~A. Efros, ``Unpaired image-to-image
  translation using cycle-consistent adversarial networks,'' in
  \emph{Proceedings of the IEEE international conference on computer vision},
  2017, pp. 2223--2232.

\bibitem{bychkovsky2011learning}
V.~Bychkovsky, S.~Paris, E.~Chan, and F.~Durand, ``Learning photographic global
  tonal adjustment with a database of input/output image pairs,'' in \emph{CVPR
  2011}.\hskip 1em plus 0.5em minus 0.4em\relax IEEE, 2011, pp. 97--104.

\bibitem{deng2009imagenet}
J.~Deng, W.~Dong, R.~Socher, L.-J. Li, K.~Li, and L.~Fei-Fei, ``Imagenet: A
  large-scale hierarchical image database,'' in \emph{2009 IEEE conference on
  computer vision and pattern recognition}.\hskip 1em plus 0.5em minus
  0.4em\relax Ieee, 2009, pp. 248--255.

\bibitem{lowe1999object}
D.~G. Lowe, ``Object recognition from local scale-invariant features,'' in
  \emph{Proceedings of the seventh IEEE international conference on computer
  vision}, vol.~2.\hskip 1em plus 0.5em minus 0.4em\relax Ieee, 1999, pp.
  1150--1157.

\bibitem{xiao2010sun}
J.~Xiao, J.~Hays, K.~A. Ehinger, A.~Oliva, and A.~Torralba, ``Sun database:
  Large-scale scene recognition from abbey to zoo,'' in \emph{2010 IEEE
  computer society conference on computer vision and pattern
  recognition}.\hskip 1em plus 0.5em minus 0.4em\relax IEEE, 2010, pp.
  3485--3492.

\bibitem{mildenhall2021nerf}
B.~Mildenhall, P.~P. Srinivasan, M.~Tancik, J.~T. Barron, R.~Ramamoorthi, and
  R.~Ng, ``Nerf: Representing scenes as neural radiance fields for view
  synthesis,'' \emph{Communications of the ACM}, vol.~65, no.~1, pp. 99--106,
  2021.

\bibitem{lee2009geometric}
D.~C. Lee, M.~Hebert, and T.~Kanade, ``Geometric reasoning for single image
  structure recovery,'' in \emph{2009 IEEE conference on computer vision and
  pattern recognition}.\hskip 1em plus 0.5em minus 0.4em\relax IEEE, 2009, pp.
  2136--2143.

\bibitem{nguyen2014raw}
R.~Nguyen, D.~K. Prasad, and M.~S. Brown, ``Raw-to-raw: Mapping between image
  sensor color responses,'' in \emph{Proceedings of the IEEE Conference on
  Computer Vision and Pattern Recognition}, 2014, pp. 3398--3405.

\bibitem{chakrabarti2009empirical}
A.~Chakrabarti, D.~Scharstein, and T.~Zickler, ``An empirical camera model for
  internet color vision.'' in \emph{BMVC}, vol.~1, no.~2.\hskip 1em plus 0.5em
  minus 0.4em\relax Citeseer, 2009, p.~4.

\bibitem{akkaynak2017space}
D.~Akkaynak, T.~Treibitz, T.~Shlesinger, Y.~Loya, R.~Tamir, and D.~Iluz, ``What
  is the space of attenuation coefficients in underwater computer vision?'' in
  \emph{Proceedings of the IEEE Conference on Computer Vision and Pattern
  Recognition}, 2017, pp. 4931--4940.

\bibitem{afifi2021semi}
M.~Afifi and A.~Abuolaim, ``Semi-supervised raw-to-raw mapping,'' \emph{arXiv
  preprint arXiv:2106.13883}, 2021.

\bibitem{cheng2014illuminant}
D.~Cheng, D.~K. Prasad, and M.~S. Brown, ``Illuminant estimation for color
  constancy: why spatial-domain methods work and the role of the color
  distribution,'' \emph{JOSA A}, vol.~31, no.~5, pp. 1049--1058, 2014.

\bibitem{bhandari2022computational}
A.~Bhandari, A.~Kadambi, and R.~Raskar, \emph{Computational Imaging}.\hskip 1em
  plus 0.5em minus 0.4em\relax MIT Press, 2022.

\bibitem{karaimer2018improving}
H.~C. Karaimer and M.~S. Brown, ``Improving color reproduction accuracy on
  cameras,'' in \emph{Proceedings of the IEEE Conference on Computer Vision and
  Pattern Recognition}, 2018, pp. 6440--6449.

\bibitem{munoz2011bssrdf}
A.~Munoz, J.~I. Echevarria, F.~J. Seron, J.~Lopez-Moreno, M.~Glencross, and
  D.~Gutierrez, ``Bssrdf estimation from single images,'' in \emph{Computer
  Graphics Forum}, vol.~30, no.~2.\hskip 1em plus 0.5em minus 0.4em\relax Wiley
  Online Library, 2011, pp. 455--464.

\bibitem{debevec2004estimating}
P.~Debevec, C.~Tchou, A.~Gardner, T.~Hawkins, C.~Poullis, J.~Stumpfel,
  A.~Jones, N.~Yun, P.~Einarsson, T.~Lundgren \emph{et~al.}, ``Estimating
  surface reflectance properties of a complex scene under captured natural
  illumination,'' UNIVERSITY OF SOUTHERN CALIFORNIA LOS ANGELES, Tech. Rep.,
  2004.

\bibitem{yu1999inverse}
Y.~Yu, P.~Debevec, J.~Malik, and T.~Hawkins, ``Inverse global illumination:
  Recovering reflectance models of real scenes from photographs,'' in
  \emph{Proceedings of the 26th annual conference on Computer graphics and
  interactive techniques}, 1999, pp. 215--224.

\bibitem{oh2016yourself}
S.~W. Oh, M.~S. Brown, M.~Pollefeys, and S.~J. Kim, ``Do it yourself
  hyperspectral imaging with everyday digital cameras,'' in \emph{Proceedings
  of the IEEE Conference on Computer Vision and Pattern Recognition}, 2016, pp.
  2461--2469.

\bibitem{galliani2017learned}
S.~Galliani, C.~Lanaras, D.~Marmanis, E.~Baltsavias, and K.~Schindler,
  ``Learned spectral super-resolution,'' \emph{arXiv preprint
  arXiv:1703.09470}, 2017.

\bibitem{barnard2002camera}
K.~Barnard and B.~Funt, ``Camera characterization for color research,''
  \emph{Color Research \& Application}, vol.~27, no.~3, pp. 152--163, 2002.

\bibitem{lo2021clcc}
Y.-C. Lo, C.-C. Chang, H.-C. Chiu, Y.-H. Huang, C.-P. Chen, Y.-L. Chang, and
  K.~Jou, ``Clcc: Contrastive learning for color constancy,'' in
  \emph{Proceedings of the IEEE/CVF Conference on Computer Vision and Pattern
  Recognition}, 2021, pp. 8053--8063.

\bibitem{el2021ntire}
M.~El~Helou, R.~Zhou, S.~Susstrunk, and R.~Timofte, ``Ntire 2021 depth guided
  image relighting challenge,'' in \emph{Proceedings of the IEEE/CVF Conference
  on Computer Vision and Pattern Recognition}, 2021, pp. 566--577.

\bibitem{afifi2021cross}
M.~Afifi, J.~T. Barron, C.~LeGendre, Y.-T. Tsai, and F.~Bleibel, ``Cross-camera
  convolutional color constancy,'' in \emph{Proceedings of the IEEE/CVF
  International Conference on Computer Vision}, 2021, pp. 1981--1990.

\bibitem{afifi2021cie}
M.~Afifi, A.~Abdelhamed, A.~Abuolaim, A.~Punnappurath, and M.~S. Brown, ``Cie
  xyz net: Unprocessing images for low-level computer vision tasks,''
  \emph{IEEE Transactions on Pattern Analysis and Machine Intelligence},
  vol.~44, no.~9, pp. 4688--4700, 2021.

\bibitem{lou2015color}
Z.~Lou, T.~Gevers, N.~Hu, M.~P. Lucassen \emph{et~al.}, ``Color constancy by
  deep learning.'' in \emph{BMVC}, 2015, pp. 76--1.

\bibitem{alsam2002recovering}
A.~Alsam and G.~Finlayson, ``Recovering spectral sensitivities with
  uncertainty,'' in \emph{Conference on Colour in Graphics, Imaging, and
  Vision}, vol. 2002, no.~1.\hskip 1em plus 0.5em minus 0.4em\relax Society for
  Imaging Science and Technology, 2002, pp. 22--26.

\bibitem{hardeberg2002spectral}
J.~Y. Hardeberg, ``On the spectral dimensionality of object colours,'' in
  \emph{Conference on Colour in Graphics, Imaging, and Vision}, vol. 2002,
  no.~1.\hskip 1em plus 0.5em minus 0.4em\relax Society for Imaging Science and
  Technology, 2002, pp. 480--485.

\bibitem{parkkinen1989characteristic}
J.~P. Parkkinen, J.~Hallikainen, and T.~Jaaskelainen, ``Characteristic spectra
  of munsell colors,'' \emph{JOSA A}, vol.~6, no.~2, pp. 318--322, 1989.

\bibitem{martinez2002calculation}
F.~M. Mart{\'\i}nez-Verd{\'u}, J.~Pujol~Ramo, P.~Capilla~Perea \emph{et~al.},
  ``Calculation of the color matching functions of digital cameras from their
  complete spectral sensitivities,'' \emph{Journal of Imaging Science and
  Technology}, vol.~46, pp. 15--25, 2002.

\bibitem{AdobeDNGSpec}
``{Adobe Digital Netagive (DNG)} specification, version 1.6.0.0,''
  \url{https://helpx.adobe.com/content/dam/help/en/photoshop/pdf/dng_spec_1_6_0_0.pdf},
  accessed: 2023-04-02.

\bibitem{darrodi2015reference}
M.~M. Darrodi, G.~Finlayson, T.~Goodman, and M.~Mackiewicz, ``Reference data
  set for camera spectral sensitivity estimation,'' \emph{JOSA A}, vol.~32,
  no.~3, pp. 381--391, 2015.

\bibitem{macdonald2002colour}
L.~MacDonald and W.~Ji, ``Colour characterisation of a high-resolution digital
  camera,'' in \emph{Conference on Colour in Graphics, Imaging, and Vision},
  vol. 2002, no.~1.\hskip 1em plus 0.5em minus 0.4em\relax Society for Imaging
  Science and Technology, 2002, pp. 433--437.

\bibitem{vora1997digital}
P.~L. Vora, J.~E. Farrell, J.~D. Tietz, and D.~H. Brainard, ``Digital color
  cameras—2—spectral response,'' 1997.

\bibitem{hubel1994comparison}
P.~M. Hubel, D.~Sherman, and J.~E. Farrell, ``A comparison of methods of sensor
  spectral sensitivity estimation,'' in \emph{Color and Imaging Conference},
  vol. 1994, no.~1.\hskip 1em plus 0.5em minus 0.4em\relax Society for Imaging
  Science and Technology, 1994, pp. 45--48.

\bibitem{hardeberg1998spectral}
J.~Y. Hardeberg, H.~Brettel, and F.~J. Schmitt, ``Spectral characterization of
  electronic cameras,'' in \emph{SYBEN-Broadband European Networks and
  Electronic Image Capture and Publishing}.\hskip 1em plus 0.5em minus
  0.4em\relax International Society for Optics and Photonics, 1998, pp.
  100--109.

\bibitem{finlayson1998recovering}
G.~D. Finlayson, S.~Hordley, and P.~M. Hubel, ``Recovering device sensitivities
  with quadratic programming,'' in \emph{Color and imaging conference}, vol.
  1998, no.~1.\hskip 1em plus 0.5em minus 0.4em\relax Society for Imaging
  Science and Technology, 1998, pp. 90--95.

\bibitem{barnard1999camera}
K.~Barnard and B.~V. Funt, ``Camera calibration for color research,'' in
  \emph{Human vision and electronic imaging IV}, vol. 3644.\hskip 1em plus
  0.5em minus 0.4em\relax SPIE, 1999, pp. 576--585.

\bibitem{dyas2000robust}
B.~Dyas, ``Robust color sensor response characterization,'' in \emph{Color and
  Imaging Conference}, vol. 2000, no.~1.\hskip 1em plus 0.5em minus 0.4em\relax
  Society for Imaging Science and Technology, 2000, pp. 144--148.

\bibitem{carvalho2001learning}
P.~Carvalho, A.~Santos, A.~Dourado, and B.~Ribeiro, ``Learning spectral
  calibration parameters for color inspection,'' in \emph{Proceedings Eighth
  IEEE International Conference on Computer Vision. ICCV 2001}, vol.~2.\hskip
  1em plus 0.5em minus 0.4em\relax IEEE, 2001, pp. 660--667.

\bibitem{quan2003comparative}
S.~Quan, N.~Ohta, and X.~Jiang, ``Comparative study on sensor spectral
  sensitivity estimation,'' in \emph{Color Imaging VIII: Processing, Hardcopy,
  and Applications}, vol. 5008.\hskip 1em plus 0.5em minus 0.4em\relax SPIE,
  2003, pp. 209--220.

\bibitem{dicarlo2004emissive}
J.~M. DiCarlo, G.~E. Montgomery, and S.~W. Trovinger, ``Emissive chart for
  imager calibration,'' in \emph{Color and Imaging Conference}, vol. 2004,
  no.~1.\hskip 1em plus 0.5em minus 0.4em\relax Society for Imaging Science and
  Technology, 2004, pp. 295--301.

\bibitem{ebner2007estimating}
M.~Ebner, ``Estimating the spectral sensitivity of a digital sensor using
  calibration targets,'' in \emph{Proceedings of the 9th annual conference on
  Genetic and evolutionary computation}, 2007, pp. 642--649.

\bibitem{mauer2009measuring}
C.~Mauer and D.~Wueller, ``Measuring the spectral response with a set of
  interference filters,'' in \emph{IS\&T/SPIE Electronic Imaging}.\hskip 1em
  plus 0.5em minus 0.4em\relax International Society for Optics and Photonics,
  2009, pp. 72\,500S--72\,500S.

\bibitem{zhao2009estimating}
H.~Zhao, R.~Kawakami, R.~T. Tan, and K.~Ikeuchi, ``Estimating basis functions
  for spectral sensitivity of digital cameras,'' in \emph{Meeting on Image
  Recognition and Understanding}, vol. 2009, no.~1, 2009.

\bibitem{rump2011practical}
M.~Rump, A.~Zinke, and R.~Klein, ``Practical spectral characterization of
  trichromatic cameras,'' in \emph{Proceedings of the 2011 SIGGRAPH Asia
  Conference}, 2011, pp. 1--10.

\bibitem{pike2011using}
T.~W. Pike, ``Using digital cameras to investigate animal colouration:
  estimating sensor sensitivity functions,'' \emph{Behavioral Ecology and
  Sociobiology}, vol.~65, pp. 849--858, 2011.

\bibitem{bedarkar2011camera}
R.~S. Bedarkar, \emph{Camera spectral sensitivity characterization using a
  blackbody source}.\hskip 1em plus 0.5em minus 0.4em\relax University of
  Maryland, College Park, 2011.

\bibitem{han2012camera}
S.~Han, Y.~Matsushita, I.~Sato, T.~Okabe, and Y.~Sato, ``Camera spectral
  sensitivity estimation from a single image under unknown illumination by
  using fluorescence,'' in \emph{Proc. IEEE CVPR}, 2012, pp. 805--812.

\bibitem{jiang2013space}
J.~Jiang, D.~Liu, J.~Gu, and S.~S{\"u}sstrunk, ``What is the space of spectral
  sensitivity functions for digital color cameras?'' in \emph{IEEE Workshop on
  Applications of Computer Vision (WACV)}, 2013, pp. 168--179.

\bibitem{prasad2013quick}
D.~Prasad, R.~Nguyen, and M.~Brown, ``Quick approximation of camera's spectral
  response from casual lighting,'' in \emph{Proc. IEEE ICCV Workshops}, 2013,
  pp. 844--851.

\bibitem{kawakami2013camera}
R.~Kawakami, H.~Zhao, R.~T. Tan, and K.~Ikeuchi, ``Camera spectral sensitivity
  and white balance estimation from sky images,'' \emph{IJCV}, vol. 105, no.~3,
  pp. 187--204, 2013.

\bibitem{bongiorno2013spectral}
D.~L. Bongiorno, M.~Bryson, D.~G. Dansereau, and S.~B. Williams, ``Spectral
  characterization of cots rgb cameras using a linear variable edge filter,''
  in \emph{Digital Photography IX}, vol. 8660.\hskip 1em plus 0.5em minus
  0.4em\relax SPIE, 2013, pp. 157--166.

\bibitem{huynh2014recovery}
C.~P. Huynh, A.~Robles-Kelly \emph{et~al.}, ``Recovery of spectral sensitivity
  functions from a colour chart image under unknown spectrally smooth
  illumination.'' in \emph{ICPR}, 2014, pp. 708--713.

\bibitem{macdonald2015determining}
L.~W. MacDonald, ``Determining camera spectral responsivity with multispectral
  transmission filters,'' in \emph{Color and Imaging Conference},
  vol.~23.\hskip 1em plus 0.5em minus 0.4em\relax Society for Imaging Science
  and Technology, 2015, pp. 12--17.

\bibitem{bartczak2015led}
P.~Bartczak, A.~Gebejes, P.~Falt, J.~Parkkinen, and P.~Silfstein, ``Led-based
  spectrally tunable light source for camera characterization,'' in \emph{2015
  Colour and Visual Computing Symposium (CVCS)}.\hskip 1em plus 0.5em minus
  0.4em\relax IEEE, 2015, pp. 1--5.

\bibitem{finlayson2016rank}
G.~Finlayson, M.~M. Darrodi, and M.~Mackiewicz, ``Rank-based camera spectral
  sensitivity estimation,'' \emph{JOSA A}, vol.~33, no.~4, pp. 589--599, 2016.

\bibitem{manakov2016evaluation}
A.~Manakov, ``Evaluation of computational radiometric and spectral sensor
  calibration techniques,'' in \emph{Optics, Photonics and Digital Technologies
  for Imaging Applications IV}, vol. 9896.\hskip 1em plus 0.5em minus
  0.4em\relax SPIE, 2016, pp. 130--143.

\bibitem{zhang2017camera}
L.~Zhang, Y.~Fu, Y.~Zheng, and H.~Huang, ``Camera spectral sensitivity,
  illumination and spectral reflectance estimation for a hybrid hyperspectral
  image capture system,'' in \emph{2017 IEEE International Conference on Image
  Processing (ICIP)}.\hskip 1em plus 0.5em minus 0.4em\relax IEEE, 2017, pp.
  545--545.

\bibitem{chaji2018estimation}
S.~Chaji, A.~Pourreza, H.~Pourreza, and M.~Rouhani, ``Estimation of the camera
  spectral sensitivity function using neural learning and architecture,''
  \emph{JOSA A}, vol.~35, no.~6, pp. 850--858, 2018.

\bibitem{karge2018using}
A.~Karge, I.~Rieger, B.~Eberhardt, and A.~Schilling, ``Using chromaticity error
  minimisation for fast camera spectral responsivity measurement,'' in
  \emph{Color and Imaging Conference}, vol. 2018, no.~1.\hskip 1em plus 0.5em
  minus 0.4em\relax Society for Imaging Science and Technology, 2018, pp.
  67--74.

\bibitem{zhu2020spectral}
J.~Zhu, X.~Xie, N.~Liao, Z.~Zhang, W.~Wu, and L.~Lv, ``Spectral sensitivity
  estimation of trichromatic camera based on orthogonal test and window
  filtering,'' \emph{Optics Express}, vol.~28, no.~19, pp. 28\,085--28\,100,
  2020.

\bibitem{toivonen2020practical}
M.~E. Toivonen and A.~Klami, ``Practical camera sensor spectral response and
  uncertainty estimation,'' \emph{Journal of Imaging}, vol.~6, no.~8, p.~79,
  2020.

\bibitem{ji2021compressive}
Y.~Ji, Y.~Kwak, S.~M. Park, and Y.~L. Kim, ``Compressive recovery of smartphone
  rgb spectral sensitivity functions,'' \emph{Optics Express}, vol.~29, no.~8,
  pp. 11\,947--11\,961, 2021.

\bibitem{tominaga2021measurement}
S.~Tominaga, S.~Nishi, and R.~Ohtera, ``Measurement and estimation of spectral
  sensitivity functions for mobile phone cameras,'' \emph{Sensors}, vol.~21,
  no.~15, p. 4985, 2021.

\bibitem{ma2021recovery}
L.~Ma, Q.~Gao, S.~Chen, C.~Li, and K.~Xiao, ``Recovery of camera spectral
  sensitivity based on multi-objective optimization method,'' in \emph{First
  Optics Frontier Conference}, vol. 11850.\hskip 1em plus 0.5em minus
  0.4em\relax SPIE, 2021, pp. 304--309.

\bibitem{xu2022rank}
B.~Xu, L.~Ma, and P.~Li, ``Rank-based camera spectral sensitivity estimation
  under multiple illuminations,'' in \emph{Second Optics Frontier Conference
  (OFS 2022)}, vol. 12307.\hskip 1em plus 0.5em minus 0.4em\relax SPIE, 2022,
  pp. 46--53.

\bibitem{fan2023camera}
H.~Fan and M.~R. Luo, ``Camera spectral sensitivity estimation based on a
  display,'' in \emph{Innovative Technologies for Printing and
  Packaging}.\hskip 1em plus 0.5em minus 0.4em\relax Springer, 2023, pp.
  24--30.

\bibitem{exiftool}
\BIBentryALTinterwordspacing
{Phil Harvey}, \emph{ExifTool}, Kingston, Ontario, Canada, 2016. [Online].
  Available: \url{https://exiftool.org/}
\BIBentrySTDinterwordspacing

\bibitem{RawSamples}
``{Rawsamples.ch},'' \url{http://rawsamples.ch/index.php/en/}, accessed:
  2023-04-11.

\bibitem{AdobeMatrices}
``{Ilia Sibiryakov's github},''
  \url{https://github.com/ilia3101/ExtractAdobeCameraMatrices}, accessed:
  2023-04-11.

\bibitem{datathief}
``{DataThief III by B. Tummers, 2006},'' \url{https://datathief.org/},
  accessed: 2023-04-11.

\bibitem{stevens2007using}
M.~Stevens, C.~A. P{\'a}rraga, I.~C. Cuthill, J.~C. Partridge, and T.~S.
  Troscianko, ``Using digital photography to study animal coloration,''
  \emph{Biological Journal of the Linnean society}, vol.~90, no.~2, pp.
  211--237, 2007.

\bibitem{akkaynak2011using}
D.~Akkaynak, E.~Chan, J.~J. Allen, and R.~T. Hanlon, ``Using spectrometry and
  photography to study color underwater,'' in \emph{Proc. MTS/IEEE OCEANS},
  2011.

\bibitem{berra2015estimation}
E.~Berra, S.~Gibson-Poole, A.~MacArthur, R.~Gaulton, and A.~Hamilton,
  ``Estimation of the spectral sensitivity functions of un-modified and
  modified commercial off-the-shelf digital cameras to enable their use as a
  multispectral imaging system for uavs,'' in \emph{International Conference on
  Unmanned Aerial Vehicles in Geomatics}.\hskip 1em plus 0.5em minus
  0.4em\relax Newcastle University, 2015.

\bibitem{brady2009camera}
M.~Brady and G.~E. Legge, ``Camera calibration for natural image studies and
  vision research,'' \emph{JOSA A}, vol.~26, no.~1, pp. 30--42, 2009.

\bibitem{huynh2007comparative}
C.~P. Huynh and A.~Robles-Kelly, ``Comparative colorimetric simulation and
  evaluation of digital cameras using spectroscopy data,'' in \emph{9th
  Biennial Conference of the Australian Pattern Recognition Society on Digital
  Image Computing Techniques and Applications (DICTA 2007)}.\hskip 1em plus
  0.5em minus 0.4em\relax IEEE, 2007, pp. 309--316.

\bibitem{lebourgeois2008can}
V.~Lebourgeois, A.~B{\'e}gu{\'e}, S.~Labb{\'e}, B.~Mallavan, L.~Pr{\'e}vot, and
  B.~Roux, ``Can commercial digital cameras be used as multispectral sensors? a
  crop monitoring test,'' \emph{Sensors}, vol.~8, no.~11, pp. 7300--7322, 2008.

\bibitem{conde2022reversed}
M.~V. Conde, R.~Timofte, Y.~Huang, J.~Peng, C.~Chen, C.~Li,
  E.~P{\'e}rez-Pellitero, F.~Song, F.~Bai, S.~Liu \emph{et~al.}, ``Reversed
  image signal processing and raw reconstruction. aim 2022 challenge report,''
  in \emph{European Conference on Computer Vision}.\hskip 1em plus 0.5em minus
  0.4em\relax Springer, 2022, pp. 3--26.

\bibitem{nguyen2014training}
R.~M. Nguyen, D.~K. Prasad, and M.~S. Brown, ``Training-based spectral
  reconstruction from a single rgb image,'' in \emph{Computer Vision--ECCV
  2014: 13th European Conference, Zurich, Switzerland, September 6-12, 2014,
  Proceedings, Part VII 13}.\hskip 1em plus 0.5em minus 0.4em\relax Springer,
  2014, pp. 186--201.

\bibitem{akhtar2014sparse}
N.~Akhtar, F.~Shafait, and A.~Mian, ``Sparse spatio-spectral representation for
  hyperspectral image super-resolution,'' in \emph{Computer Vision--ECCV 2014:
  13th European Conference, Zurich, Switzerland, September 6-12, 2014,
  Proceedings, Part VII 13}.\hskip 1em plus 0.5em minus 0.4em\relax Springer,
  2014, pp. 63--78.

\bibitem{akhtar2016hierarchical}
------, ``Hierarchical beta process with gaussian process prior for
  hyperspectral image super resolution,'' in \emph{Computer Vision--ECCV 2016:
  14th European Conference, Amsterdam, The Netherlands, October 11-14, 2016,
  Proceedings, Part III 14}.\hskip 1em plus 0.5em minus 0.4em\relax Springer,
  2016, pp. 103--120.

\bibitem{luo2021time}
X.~Luo, X.~Zhang, P.~Yoo, R.~Martin-Brualla, J.~Lawrence, and S.~M. Seitz,
  ``Time-travel rephotography,'' \emph{ACM Transactions on Graphics (TOG)},
  vol.~40, no.~6, pp. 1--12, 2021.

\bibitem{karaimer2016software}
H.~C. Karaimer and M.~S. Brown, ``A software platform for manipulating the
  camera imaging pipeline,'' in \emph{Computer Vision--ECCV 2016: 14th European
  Conference, Amsterdam, The Netherlands, October 11--14, 2016, Proceedings,
  Part I 14}.\hskip 1em plus 0.5em minus 0.4em\relax Springer, 2016, pp.
  429--444.

\end{thebibliography}

\ifpeerreview \else

\vspace{-1cm}
\begin{IEEEbiography}[{\includegraphics[width=1in,height=1.25in,clip,keepaspectratio,trim=0.5cm 0cm 0.5cm 0cm]{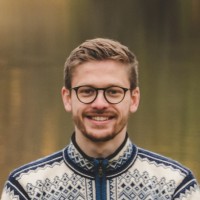}}]{Grigory Solomatov}
  is a postdoctoral fellow in the Department of Marine Technologies at the University of Haifa and is part of the COLOR Lab at the Interuniversity Institute of Marine Sciences in Eilat. He earned his PhD in mathematics and computer science from the Technical University of Denmark. His current research pivots towards leveraging techniques from optimization and machine learning to advance underwater computer vision and computational photography.
\end{IEEEbiography}

\vspace{-1cm}
\begin{IEEEbiography}
[{\includegraphics[width=1in,height=1.25in,clip,keepaspectratio]{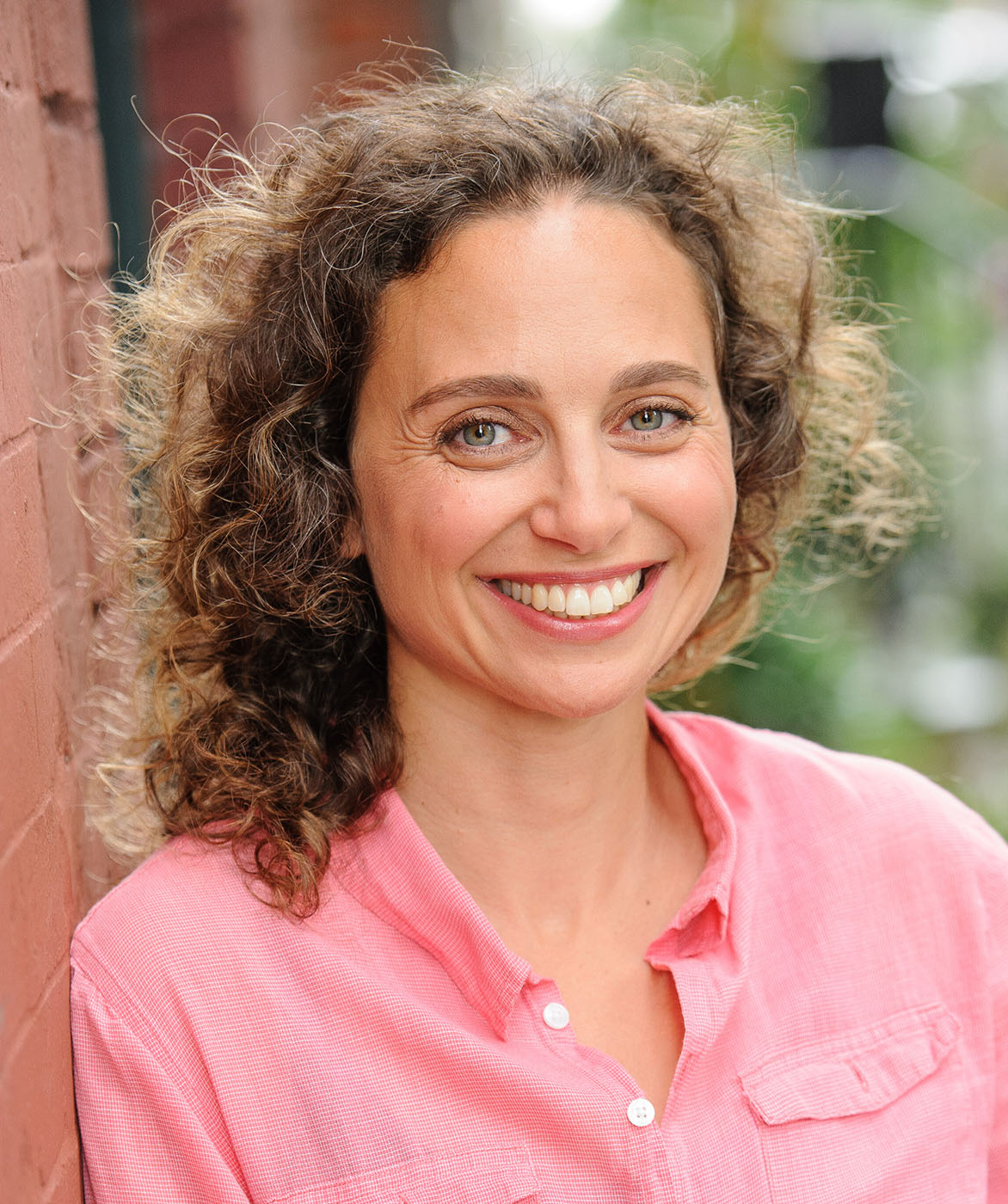}}]{Derya Akkaynak}
is an assistant professor in the Department of Marine Technologies at the University of Haifa, where she directs the COLOR Lab located at the Interuniversity Institute of Marine Sciences in Eilat. She received her PhD in Mechanical \& Oceanographic Engineering from MIT and Woods Hole Oceanographic Institution, MSc in Aeronautics and Astronautics from MIT, and BSc in Aerospace Engineering from METU (Turkey). She studies light and vision in the ocean, at the intersection of which lies color.

\end{IEEEbiography}




\fi

\end{document}


\maketitle

 \section{Details for the generation of results in \textbf{Fig.1}}\label{sec:fig1-results}
 Below, we describe how the results in Fig. 1 of main text were generated using camera spectral sensitivities.  
 
 In all examples, the camera RAW images were processed through the Adobe DNG Converter to obtain DNG files, and these were then run through steps 1-4 of the digital camera processing pipeline of Karaimer and Brown~\cite{karaimer2016software}, which allows users access into the Adobe DNG SDK. This process results in a demosaicked linear tiff image with no additional modifications. The average reflectance spectra of the ColorChecker patches were obtained from \href{https://babelcolor.com/}{Babel Colour} and the DGK Kolor Kard patches were obtained from~\cite{akkaynak2017space}. 

 The error between actual and predicted \texttt{RGB} values were calculated using the angular error:

 \begin{equation}
     err = cos^{-1} \bigg( \frac{RGB^{actual} \cdot RGB^{predicted}}{\norm{RGB^{actual}}\norm{RGB^{predicted}}}\bigg)
 \end{equation}

and this is used as the objective function for parts \textbf{A} and \textbf{B}.
 \subsection{\textbf{Fig.1 A}}
The image used is Canon600D\_0195.CR2 from the \href{https://cvil.eecs.yorku.ca/projects/public_html/illuminant/illuminant.html} {NUS dataset} of Cheng et al.~\cite{cheng2014illuminant}. The camera used was a Nikon D40, whose spectral response is made available by Jiang et al.~\cite{jiang2013space}. We discretize this spectral response into $n=31$ values in the range 400-700 nm, in steps of 10nm.

This dataset also provides the coordinates of the ColorChecker and all its patches, which we used to extract the \texttt{RGB} values of each patch from the linear tiff image. We modeled daylight by the CIE Daylight Model, which is parameterized only by the correlated color temperature (CCT). We formulated an optimization problem in which we minimized the angular \texttt{RGB} between the $m=24$ actual \texttt{RGB} values (extracted from the ColorChecker) and the predicted \texttt{RGB} values error (Eq. 1 above). The predicted \texttt{RGB} values were obtained using Eq. (1) in main text, with the light spectrum obtained from the CIE daylight model, with the only unknown being CCT. The optimization was implemented in Matlab, using the `fmincon' function, using a lower bound of 4,000K and an upper bound of 25,000K for CCT. The resulting temperature found for the example shown in Fig. 1A was 5,554 K (approximately CIE D55). The precise optimization problem is
\begin{equation*}
  \argmin_{4000\text{K} \le T \le 25000\text{K}} \norm{
    [\angle(\mat{I}_d^{[k,*]}, (\mat{R}\mat{L}_T\mat{S}_d)^{[k,*]})]_{1 \le k \le m}
  }_2 \ ,
\end{equation*}
where the superscript $[k,*]$ means taking the $k$-th row and  $\mat{L}_T \in \RR^{n \times n}$ is a diagonal illuminant matrix whose diagonal is obtained by discretizing the CIE illuminant series D, i.e.
\begin{align*}
  \label{eq:daylight-model}
  L_{T}(\lambda) &= L_0(\lambda) + M_1(T)L_1(\lambda) + M_2(T)L_2(\lambda) \ , \text{ where} \\
  M_1 &= (-1.3515-1.7703x_D(T)+5.9114y_D(T))/M \ , \\
  M_2 &= (0.0300-31.4424x_D(T)+30.0717y_D(T))/M \ , \\
  M &= 0.0241+0.2562x_D(T)-0.7341y_D(T) \ , \\
  y_D(T) &= -3.000x_D(T)^2 + 2.879x_D(T) - 0.275 , \\
  x_D(T) &= \begin{cases}
              0.244063 + 0.09911\frac{10^3}{T} + 2.9678\frac{10^6}{T^2} - 4.6070\frac{10^9}{T^3} \\
              \hspace{100pt} \text{if } 4000\text{K} \le T \le 7000\text{K} \vspace{5pt}\\
              0.237040 + 0.24748\frac{10^3}{T} + 1.9018\frac{10^6}{T^2} - 2.0064\frac{10^9}{T^3}\\    
              \hspace{100pt} \text{if } 7000\text{K} < T \le 25000\text{K} \\
            \end{cases} \ .
\end{align*}

 \subsection{\textbf{Fig.1 B}}
 The image used is \_DSC0098.tif from the dataset of Akkaynak et al.~\cite{akkaynak2017space}. The camera used was a Nikon D90, whose spectral response is also made available by Jiang et al.~\cite{jiang2013space}.
 
 Note that the color chart used here is not the Macbeth ColorChecker, but the waterproof DGK KolorKard. The $m=18$ color chart patches were manually extracted and the corresponding \texttt{RGB} values were extracted from the linear tiff image. 
 
 Then, the image formation from Eq. (1) of main text was used with the observation that the ambient light will be exponentially attenuated, i.e., 
 
 \begin{equation}
      L_{\text{depth}}(\lambda) = L_{\text{surface}}(\lambda)e^{-K_D(\lambda) \times y},
 \end{equation}
where $y$ is depth and $K_D(\lambda)$ is the diffuse downwelling attenuation coefficient, which is the quantity we would like to estimate. The depth gauge in the photo shows $y=16.2$ meters.

For surface light, it is generally safe to assume any broadband light (i.e., any CIE D-series light), as the water will attenuate all light to the same monochromatic spectrum, so here we assumed CIE D65.  Again using Eq. (1) from main text to calculate the predicted \texttt{RGB} values, we minimize the angular error between predicted and observed \texttt{RGB} values using the `fmincon' function in Matlab using a lower bound of 0 and an upper bound of 1 $m^{-1}$ for $K_D(\lambda)$. 
 
It is important to note here that $K_D(\lambda)$ cannot be solved for using 31 unknowns because this color chart has much fewer linearly independent patches. By trial and error, we ended up using discretization resolution $\widehat{n} = 10$ to approximate $K_D(\lambda)$ by a linear interpolation $\widehat{K}_D(\lambda)$ of a vector $\vec{K} \in \RR^{\widehat{n}}$. The precise optimization problem is

\begin{equation}
  \argmin_{\substack{
      \vec{K} \in \RR^{\widehat{n}} \\
      \min(\vec{K}) \ge 0 \\
      \max(\vec{K}) \le 1
    }   
  } \norm{
    [\angle(\mat{I}_d^{[k,*]}, (\mat{R}\mat{L}_{\vec{K}}\mat{S}_d)^{[k,*]})]_{1 \le k \le m}
  }_2 \ ,
\end{equation}
where $\mat{L}_{\vec{K}} \in \RR^{n \times n}$ is the diagonal matrix whose diagonal is obtained by discretizing $L_{\text{D65}}(\lambda)e^{-\widehat{K}_D(\lambda) \times y}$, where $\widehat{K}_D(\lambda)$ is the linear interpolation of $\vec{K}$.

\subsection{\textbf{Fig.1 C}}
 The cameras used in this example were a Nikon D40, and Canon600D, whose spectral responses were both made available by Jiang et al.~\cite{jiang2013space}.
 
 To do the raw-to-raw mapping example here, we followed the illumination-invariant method of~\cite{nguyen2014raw}. We skipped the pairwise calibration step, since we do not have physical access to the cameras. Instead, we used the spectral responses of both cameras, and published spectral power distributions of 23 standard CIE illuminants (namely CIE F1-12, A,B,C, D40-D75 illuminants) to compute a global mapping between the two cameras as described in~\cite{nguyen2014raw}, using Macbeth ColorChecker reflectances. 
 
 Next, we obtained the ``white-balanced' mapping by white balancing the ColorChecker values for each illuminant using the white patch of the color chart (even though the white Macbeth patch does not have a perfectly flat reflectance spectrum). 

  We then picked example images from the \href{https://cvil.eecs.yorku.ca/projects/public_html/illuminant/illuminant.html} {NUS dataset} of Cheng et al.~\cite{cheng2014illuminant}. The Canon image is Canon600D\_0091.CR2, and the Nikon image is NikonD40\_0004.NEF. Using gray-world algorithm, we estimated the illuminant in each image and white balanced the source image.
  
  Next, we applied the white-balanced global mapping to the source image, to obtain its projection into the target camera space. Then, we needed to white balance this image to the correct value in the target camera's color space; so we used the global transform to obtain the correct white point, and inverse-white balance the transformed image. We then completed the in-camera processing using the pipeline software of~\cite{karaimer2016software} to obtain the photofinished sRGB image.
  
\subsection{\textbf{Fig.1 D}}
The camera used in this example was a Nikon D40 whose spectral response was  made available by Jiang et al.~\cite{jiang2013space}. The images and the corresponding ground-truth illumination data are from the \href{https://cvil.eecs.yorku.ca/projects/public_html/illuminant/illuminant.html} {NUS dataset} of Cheng et al.~\cite{cheng2014illuminant}. There were 117 images for the NikonD40 folder of this dataset. We manually scored whether each photo was taken indoors or outdoors. 

Next, we calculated the ``white-point' of every daylight illuminant between 4,000 and 25,000 K, using the CIE daylight model, in the color space of the Nikon D40 camera. We then plotted the locus of ``daylight chromaticities" for this camera, where chromaticity $r$ is given as:

 \begin{equation}
     r = \frac{R}{(R+G+B)},
 \end{equation}

and the $b$ chromaticity is given similarly. Finally, we computed the $r$ and $b$ chromaticities of the provided ground-truth illuminations plotted them in the same coordinates.

\section{Results on our ground-truth dataset}

\begin{figure*}[h]
\centering
\includegraphics[width = 1.7\columnwidth]{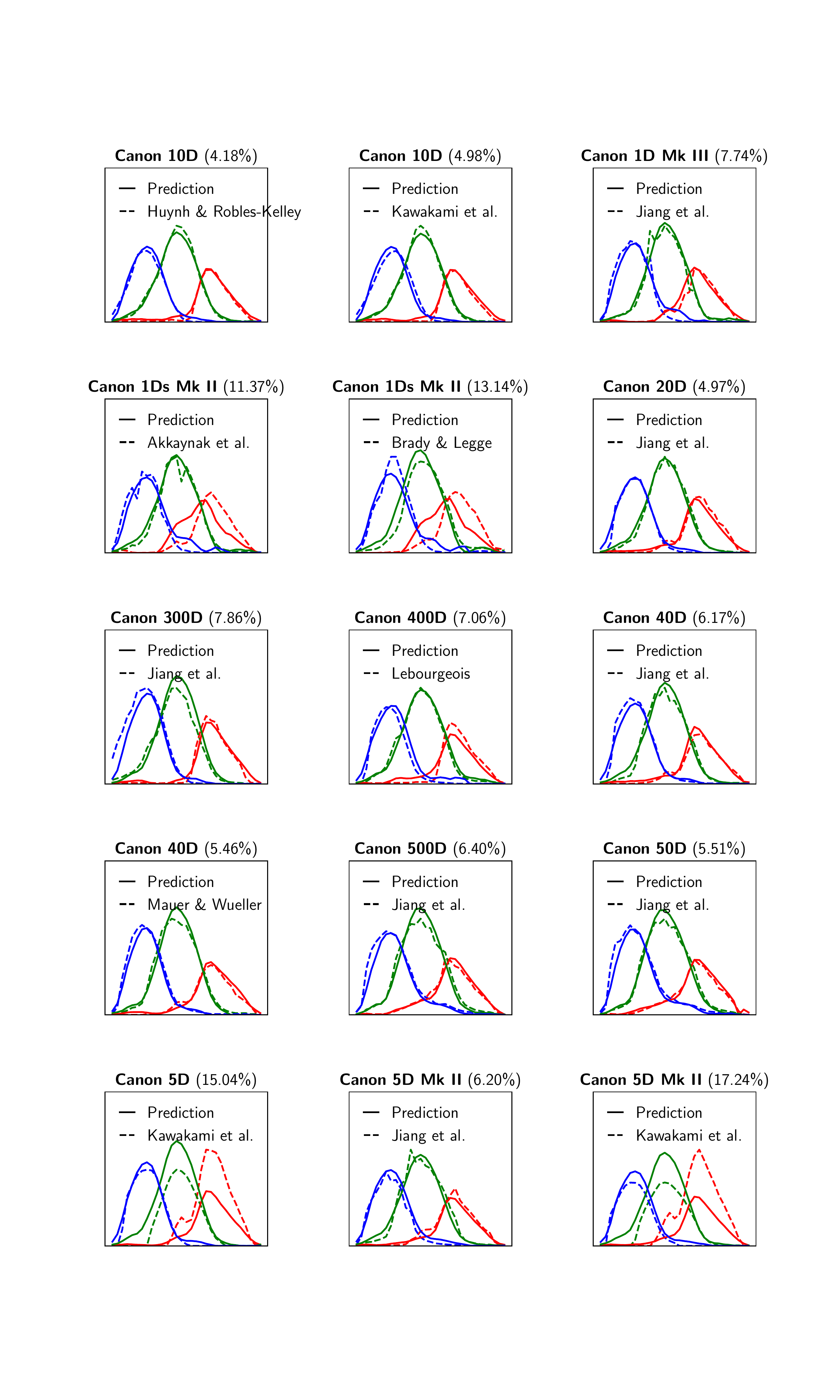}
\caption{\label{fig:fig-supp1} The results of our method on the ground-truth dataset (Part 1).}
\end{figure*}

\begin{figure*}[h]
\centering
\includegraphics[width = 1.7\columnwidth]{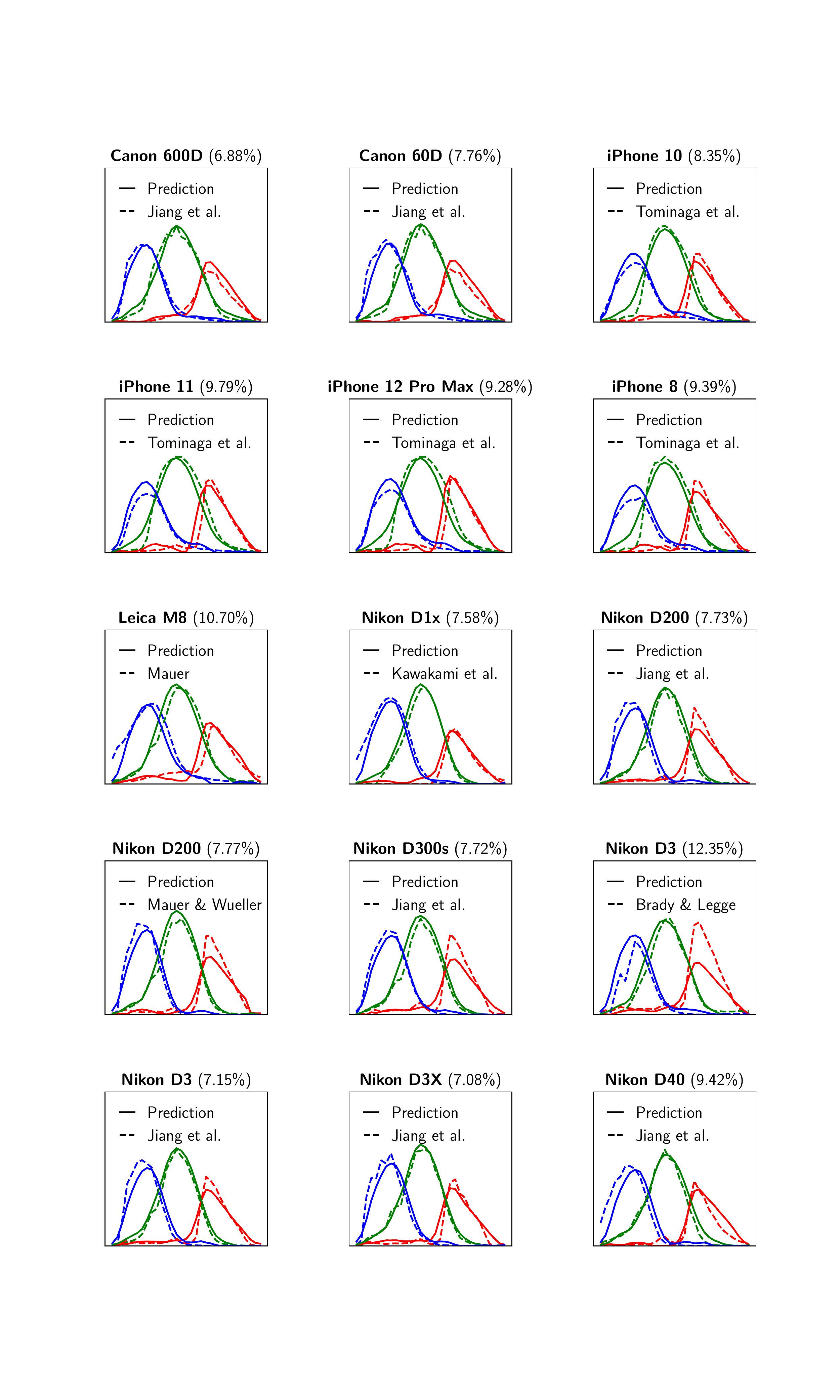}
\caption{\label{fig:fig-supp2} The results of our method on the ground-truth dataset (Part 2).}
\end{figure*}

\begin{figure*}[h]
\centering
\includegraphics[width = 1.7\columnwidth]{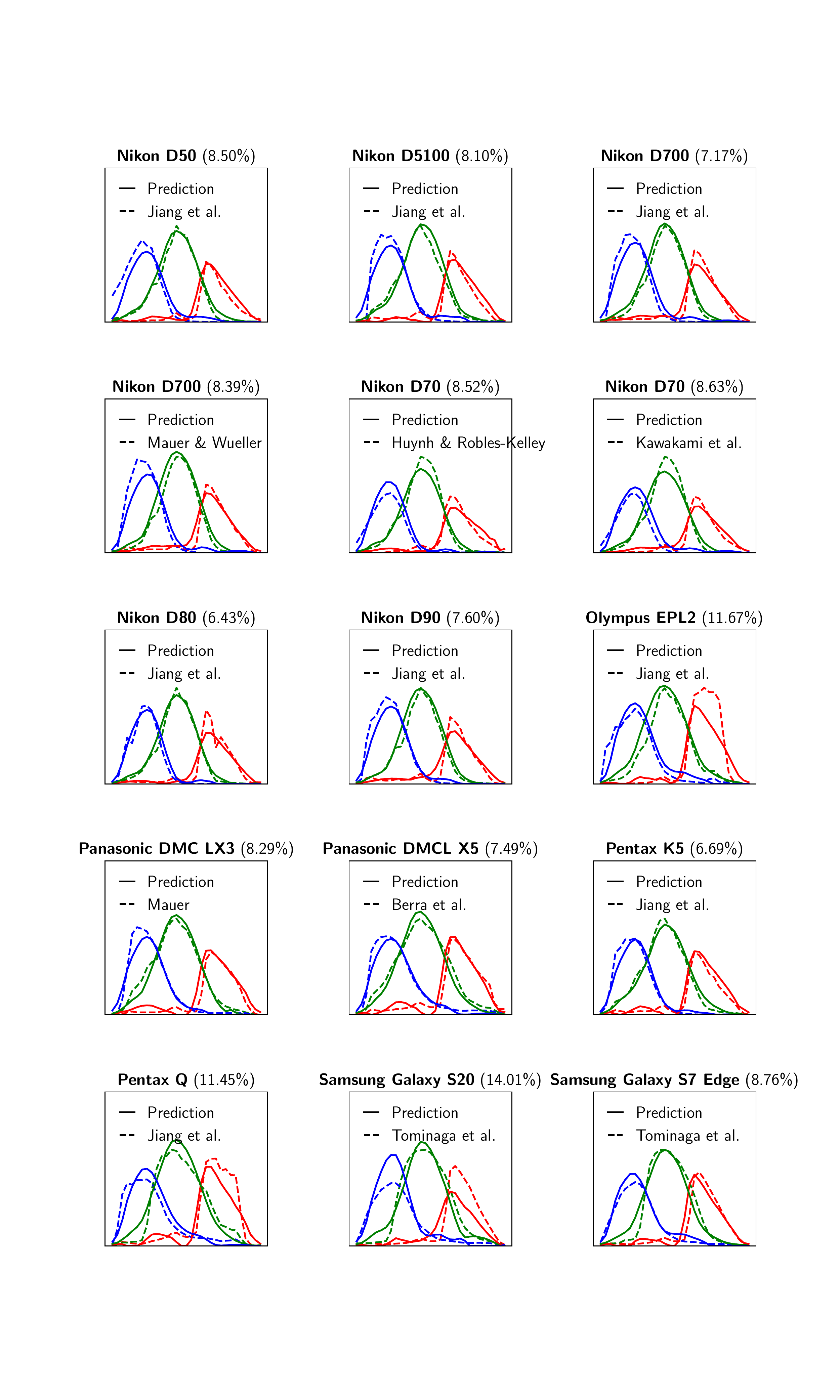}
\caption{\label{fig:fig-supp3} The results of our method on the ground-truth dataset (Part 3).}
\end{figure*}

\begin{figure*}[h]
\centering
\includegraphics[width = 2\columnwidth]{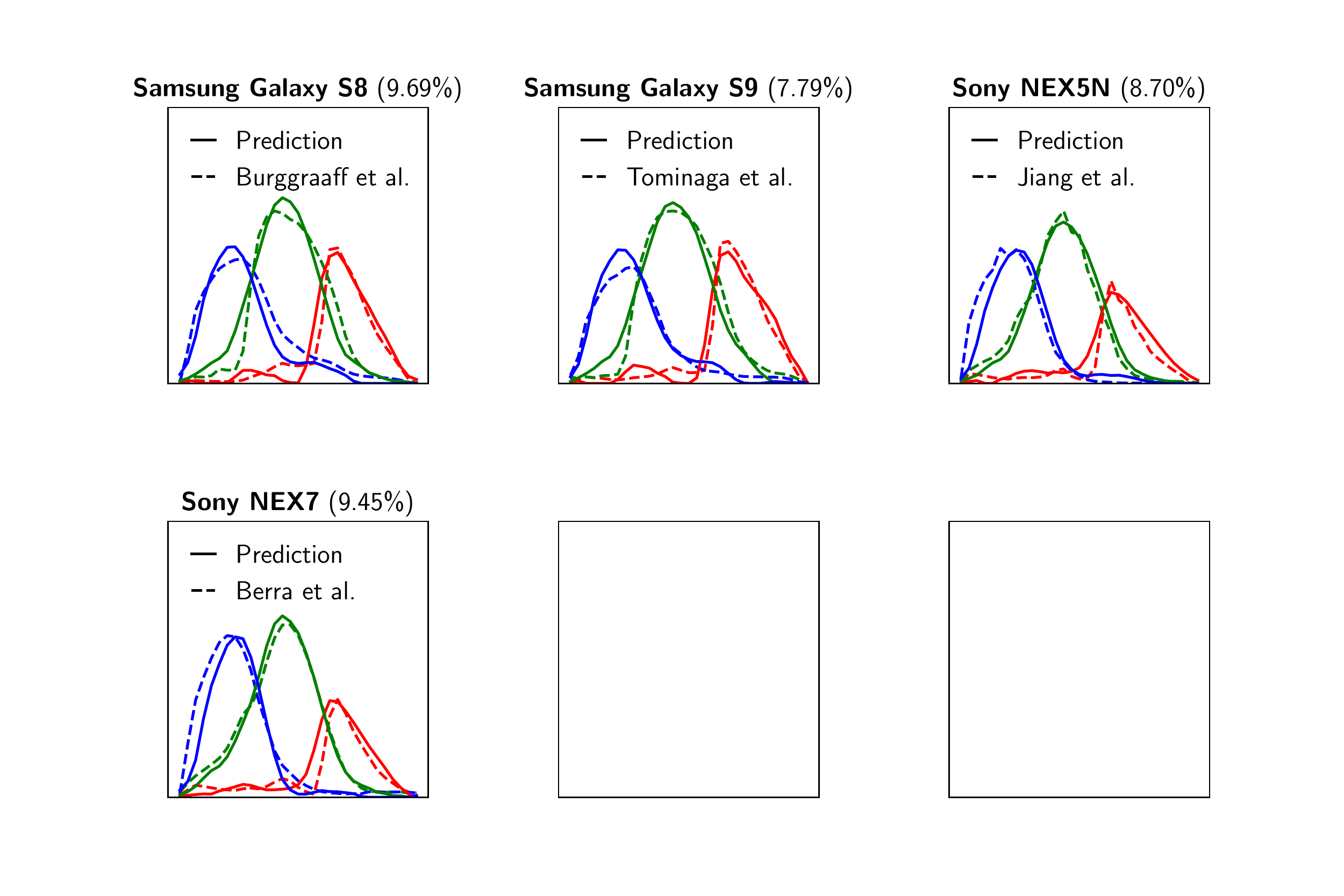}
\caption{\label{fig:fig-supp4} The results of our method on the ground-truth dataset (Part 4).}
\end{figure*}

Here, we present all the results on our ground-truth dataset in \cref{fig:fig-supp1} through \cref{fig:fig-supp4}. \change{As in Fig. 3 and Fig. 5 from the main paper, the horizontal and the vertical axes are respectively wavelength (nm) and relative sensitivity.}

\section{Relative full scale error versus $\Delta E2000$}
\begin{figure*}[h]
\centering
\includegraphics[width = 2\columnwidth]{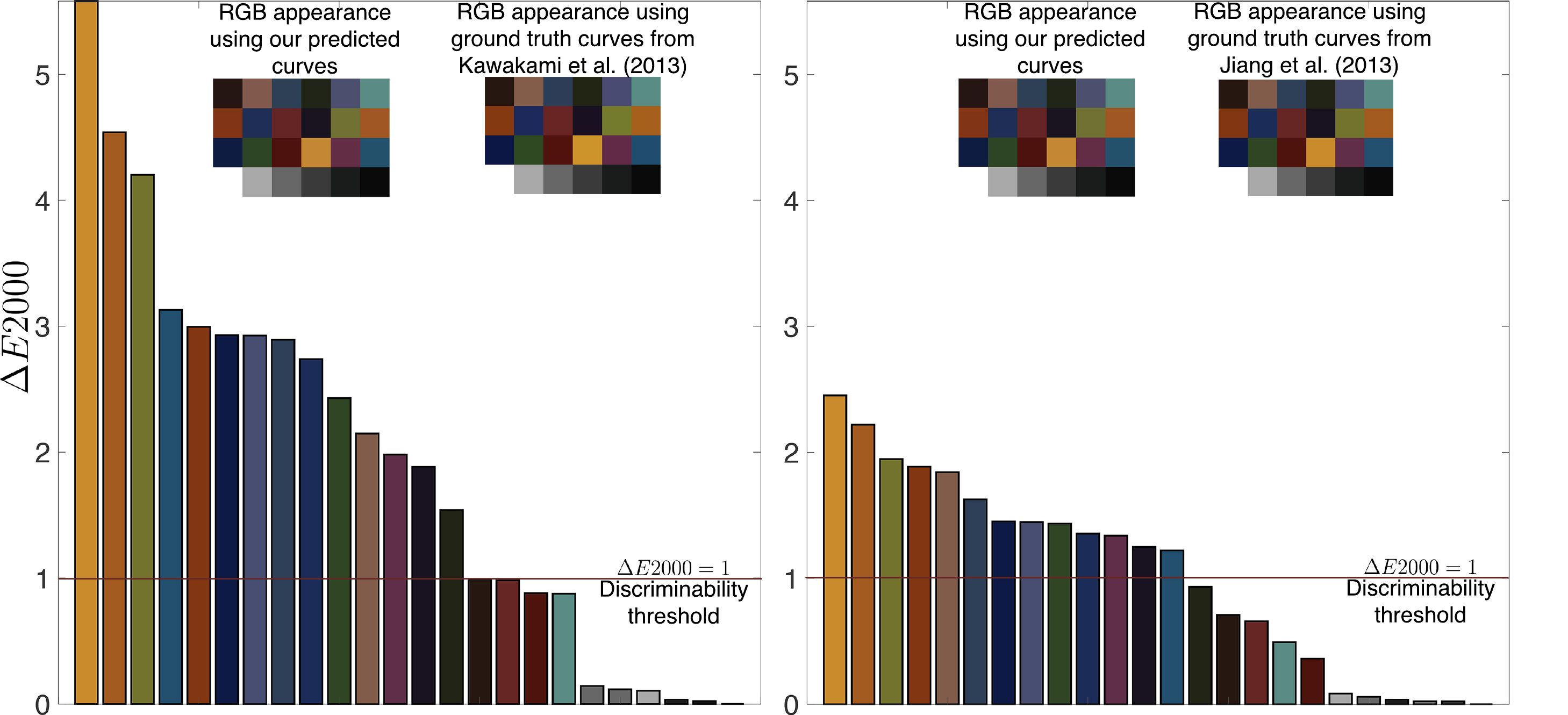}
\caption{\label{fig:de2000} Perceptual ($\Delta E2000$) errors between high-error (Kawakami et. al ground truth) and low-error (Jiang et al. ground truth) spectral sensitivity reconstructions. While the differences in the simulated color charts are hardly visible to the eye, the perceptual differences can be quantified.}
\end{figure*}

In the main text, we calculated reconstruction errors between our predicted curves and the ground truth using the relative full scale error metric. Here, using the CIE $\Delta E2000$ metric, we demonstrate the perceptual projection of these errors relative to the human visual system. We use the case of the Canon 5D Mk II, for which two ground truth values exist (Fig. 3 in main text). We simulated the appearance of a Macbeth ColorChecker under D65 light according to the Kawakami et al~\cite{kawakami2013camera} ground truth spectral sensitivity curves for the Canon 5D Mk II, and Jiang et al.~\cite{jiang2013space} ground truth. We white balanced the resulting simulations using the white patch. The distributions of the $\Delta E2000$ error resulting from the comparison between our predictions and the two ground truth curves are given in~\cref{fig:de2000}.

The color differences arising from differences in spectral sensitivity curves are difficult to discern when looking at the simulated charts, however they are reflected into the $\Delta E2000$ errors. Considering that $\Delta E2000 = 1$ marks the discriminability threshold below which two solid colors are indiscriminable to the human eye, as expected, large errors in spectral sensitivity reconstruction translates to large color differences that would be visible.

\section{A big picture look at the Adobe DNG Converter dataset}

Here, we show a breakdown of the statistics of the ~1000 camera for which we were able to extract color matrices from the Adobe DNG Converter. The information directly available from the Adobe DNG Converter were the manufacturer name, and the make \& model of the camera. We manually compiled the rest of the sensor information including sensor size, format, camera type and release year. While we made every effort to ensure the accuracy of these data and performed checks and validations, some erroneous entries may still be present.

\begin{figure*}[h]
\centering
\includegraphics[width = 2\columnwidth]{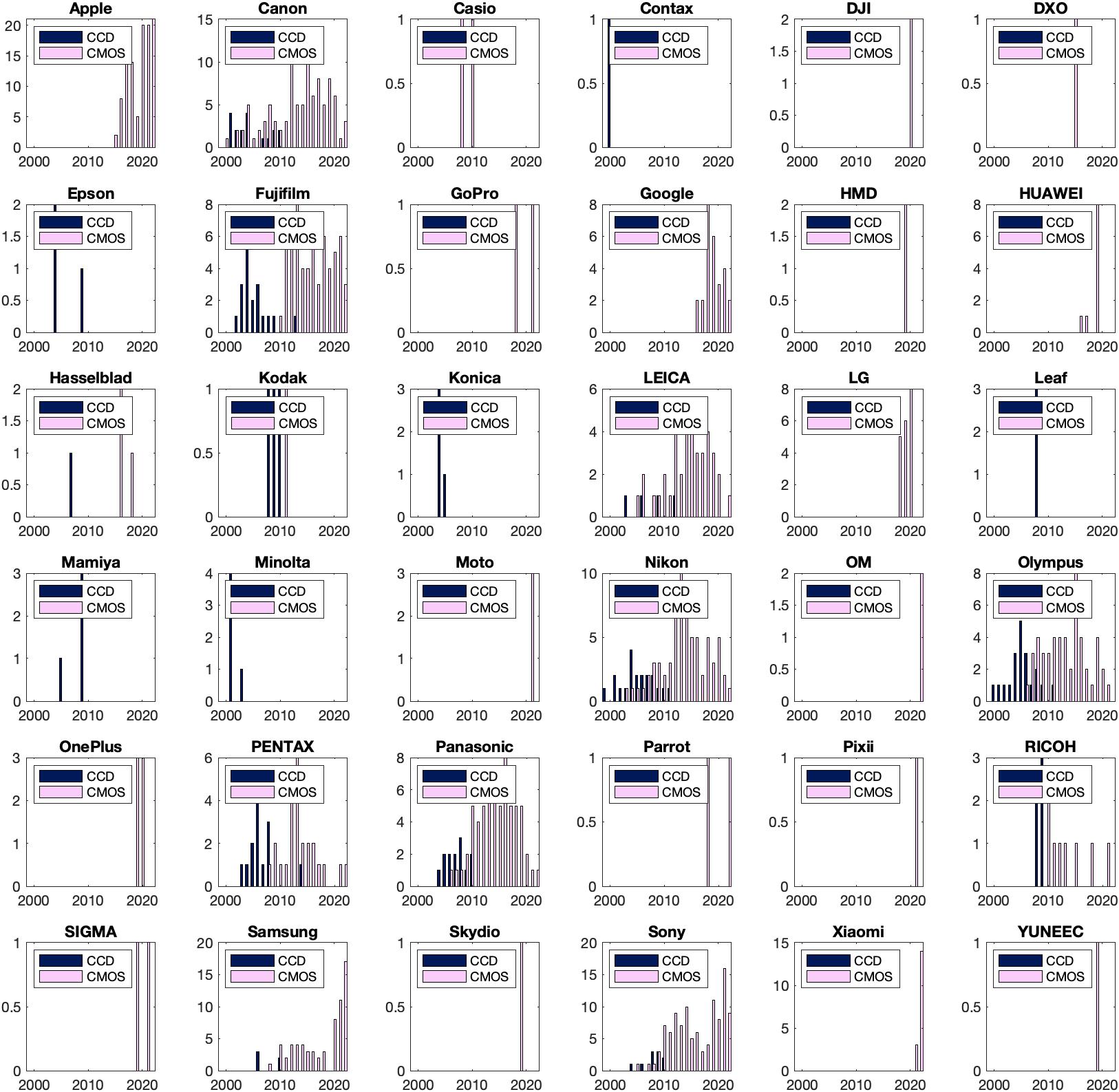}
\caption{\label{fig:sensors} Sensor types of the cameras in the Adobe dataset.}
\end{figure*}

\begin{figure*}[h]
\centering
\includegraphics[width = 2\columnwidth]{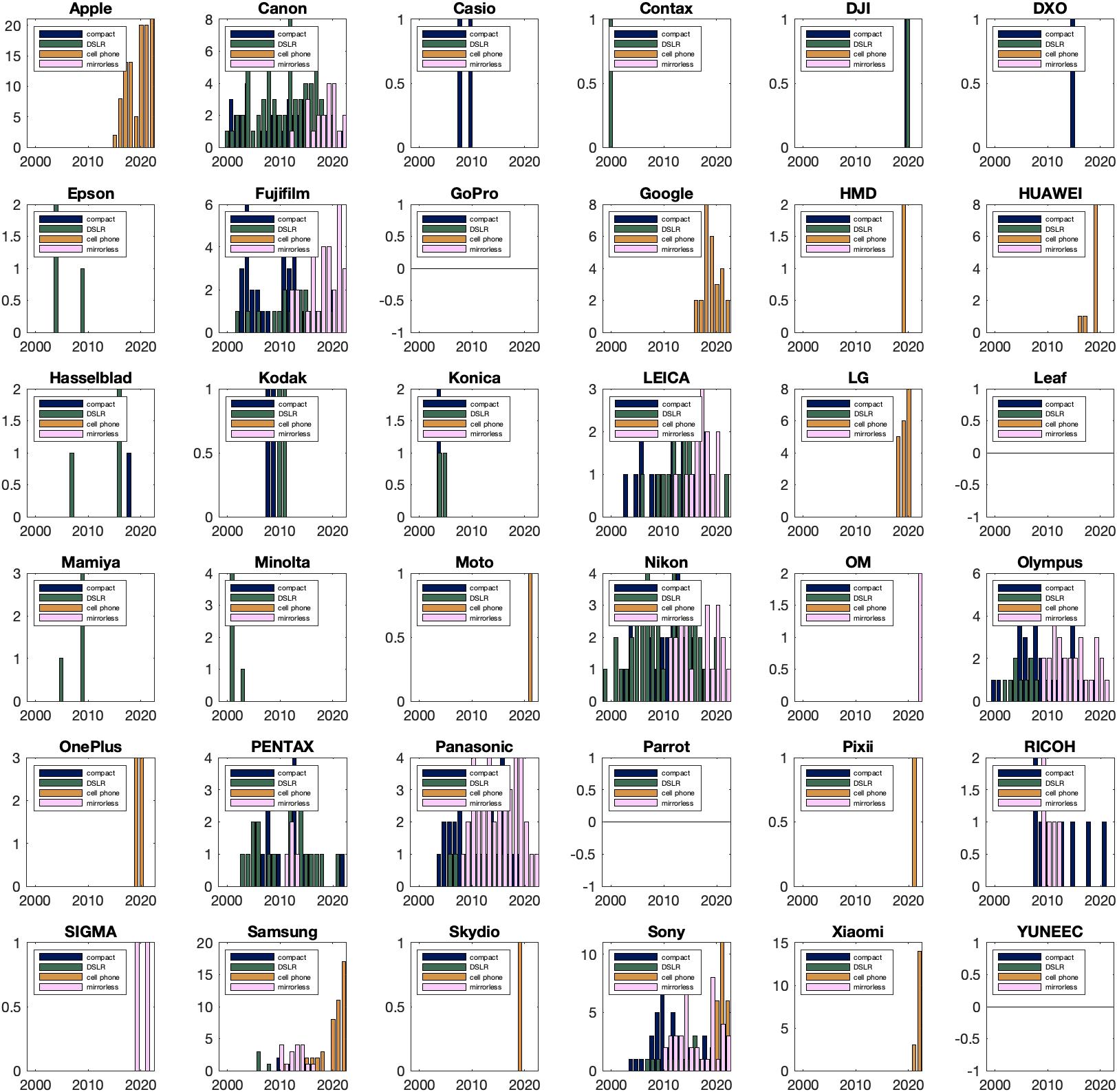}
\caption{\label{fig:cameratypes} Types of the cameras in the Adobe dataset according to release year of the camera.}
\end{figure*}

\bibliographystyle{IEEEtran}
\bibliography{references}

\ifpeerreview \else





\fi